\documentclass[fleqn,usenatbib]{mnras}

\usepackage{newtxtext,newtxmath}
\usepackage{subcaption}

\usepackage[T1]{fontenc}

\DeclareRobustCommand{\VAN}[3]{#2}
\let\VANthebibliography\thebibliography
\def\thebibliography{\DeclareRobustCommand{\VAN}[3]{##3}\VANthebibliography}

\usepackage{graphicx}	
\usepackage{amsmath}	
\usepackage{multirow}
\usepackage{ragged2e}
\usepackage{pdflscape}
\usepackage[flushleft]{threeparttable}
\usepackage{caption}
\usepackage{orcidlink}

\newcommand{\cii}{[C\,{\sc ii}]}

\newcommand{\nii}{[N\,{\sc ii}]}

\newcommand{\oiii}{[O\,{\sc iii}]}

\newcommand{\oiiil}{[O\,{\sc iii}] 88\,$\mu{\rm m}$}
\newcommand{\ciil}{[C\,{\sc ii}] 158\,$\mu{\rm m}$}

\title[Hot Dust in the Big Three Dragons]{Breathing Fire: Hot Dust in the Big Three Dragons at $z=7.15$}

\author[Rajulal et al.]{%
Gitanjali Rajulal$^{1}$\thanks{E-mail: gitanjali.rajulal13@gmail.com},
Hiddo S. B. Algera$^{1}$\thanks{E-mail: hsbalgera@asiaa.sinica.edu.tw},
Yuma Sugahara\,\orcidlink{0000-0001-6958-7856}$^{2,3}$,
Tom J. L. C. Bakx$^{4}$,
Takuya Hashimoto$^{5,6}$,
\newauthor
Suzuka Arai$^{3}$,
Akio K. Inoue$^{2,3}$,
Ikki Mitsuhashi$^{7}$,
Manuel Aravena$^{8,9}$,
Karin Cescon$^{10}$,
Chian-Chou Chen$^{1}$,
\newauthor
Elisabete da Cunha$^{11}$,
Pratika Dayal$^{12,13,14}$,
Ilse De Looze$^{15}$, 
Andreas Faisst$^{16}$,
Yoshinobu Fudamoto\,\orcidlink{0000-0001-7440-8832}$^{17}$,
\newauthor
Rodrigo Herrera-Camus$^{18,9}$, 
Hanae Inami$^{19}$,
Anton M. Koekemoer\,\orcidlink{0000-0002-6610-2048}$^{20}$,
Shoichiro Mizukoshi$^{1}$,
\newauthor
Michael Romano$^{21,22}$,
Lucie Rowland$^{10}$,
Sander Schouws$^{23,10}$, 
Renske Smit$^{24}$, 
Livia Vallini$^{25}$,
\newauthor
Wei-Hao Wang$^{1}$,
Giovanni Zamorani$^{25}$,
and Anita Zanella$^{25}$ \\
$^{1}$Institute of Astronomy and Astrophysics, Academia Sinica, 11F of Astronomy-Mathematics Building, No.1, Sec. 4, Roosevelt Rd, Taipei 106319, Taiwan, R.O.C. \\
$^{2}$Waseda Research Institute for Science and Engineering, Faculty of Science and Engineering, Waseda University, 3-4-1 Okubo, Shinjuku, Tokyo 169-8555, Japan \\
$^{3}$Department of Physics, School of Advanced Science and Engineering, Faculty of Science and Engineering, Waseda University, 3-4-1 Okubo, Shinjuku, \\ Tokyo 169-8555, Japan \\
$^{4}$Department of Space, Earth and Environment, Chalmers University of Technology, SE-412 96 Gothenburg, Sweden \\
$^{5}$Division of Physics, Faculty of Pure and Applied Sciences, University of Tsukuba, Tsukuba, Ibaraki 305-8571, Japan \\
$^6$Tomonaga Center for the History of the Universe (TCHoU), Faculty of Pure and Applied Sciences, University of Tsukuba, Tsukuba, Ibaraki 305-8571, Japan \\
$^{7}$Department for Astrophysical \& Planetary Science, University of Colorado, Boulder, CO 80309, USA \\
$^{8}$Instituto de Estudios Astrof\'{\i}sicos, Facultad de Ingenier\'{\i}a y Ciencias, Universidad Diego Portales, Av. Ej\'ercito 441, Santiago, Chile \\
$^{9}$Millennium Nucleus for Galaxies (MINGAL) \\
$^{10}$Leiden Observatory, Leiden University, P.O. Box 9513, 2300 RA Leiden, The Netherlands \\
$^{11}$International Centre for Radio Astronomy Research, University of Western Australia, 35 Stirling Hwy., Crawley, WA 6009, Australia \\
$^{12}$Canadian Institute for Theoretical Astrophysics, 60 St George St, University of Toronto, Toronto, ON M5S 3H8, Canada \\
$^{13}$David A. Dunlap Department of Astronomy and Astrophysics, University of Toronto, 50 St George St, Toronto ON M5S 3H4, Canada \\
$^{14}$Department of Physics, 60 St George St, University of Toronto, Toronto, ON M5S 3H8, Canada \\
$^{15}$Sterrenkundig Observatorium, Ghent University, Krijgslaan 281 - S9, 9000 Gent, Belgium \\
$^{16}$IPAC, California Institute of Technology, 1200 E. California Blvd., Pasadena, CA 91125, USA \\
$^{17}$Center for Frontier Science, Chiba University, 1-33 Yayoi-cho, Inage-ku, Chiba 263-8522, Japan \\
$^{18}$Departamento de Astronomía, Universidad de Concepción, Barrio Universitario, Concepción, Chile \\
$^{19}$Hiroshima Astrophysical Science Center, Hiroshima University, 1-3-1 Kagamiyama, Higashi-Hiroshima, Hiroshima 739-8526, Japan \\
$^{20}$Space Telescope Science Institute, 3700 San Martin Drive, Baltimore, MD 21218, USA \\
$^{21}$Max-Planck-Institut für Radioastronomie, Auf dem Hügel 69, D-53121, Bonn, Germany \\
$^{22}$INAF -- Osservatorio Astronomico di Padova, Vicolo dell’Osservatorio 5, I-35122 Padova, Italy \\
$^{23}$Departament d’Astronomia i Astrofìsica, Universitat de València, C. Dr. Moliner 50, E-46100 Burjassot \\
$^{24}$Astrophysics Research Institute, Liverpool John Moores University, 146 Brownlow Hill, Liverpool L3 5RF, UK \\
$^{25}$INAF -- Osservatorio di Astrofisica e Scienza dello Spazio di Bologna, Via Gobetti 93/3, 40129 Bologna, Italy
}

\date{Accepted XXX. Received YYY; in original form ZZZ}

\pubyear{2026}
\begin{document}
\label{firstpage}
\pagerange{\pageref{firstpage}--\pageref{lastpage}}
\maketitle

\begin{abstract}
We present new Atacama Large Millimeter/submillimeter Array (ALMA) Band 9 ($\lambda_\mathrm{obs} = 0.45\,\mathrm{mm}$) and 4 ($\lambda_\mathrm{obs} = 2.2\,\mathrm{mm}$) observations towards the Big Three Dragons, a pair of merging Lyman-break galaxies (LBGs) at $z=7.15$. The system was previously detected in dust continuum emission in Bands 6, 7 and 8 ($\lambda_\mathrm{obs} = 0.73 - 1.32\,\mathrm{mm}$), which, combined with our new observations, allows us to more robustly constrain its dust temperature and obscured SFR. The unresolved Band 4 observations yield a $3.5\sigma$ detection, and the $0.4''$ Band 9 observations detect the Eastern and Western LBGs at $3.7$ and $3.5\sigma$, respectively. Through optically thin modified blackbody fitting, we infer a global dust temperature of $T_d = 78_{-23}^{+35}\,\mathrm{K}$ for the system, which implies a high IR luminosity of $\log(L_\mathrm{IR}/L_\odot) = 12.32_{-0.41}^{+0.43}$. This makes the Big Three Dragons one of the most IR-luminous systems known at $z>7$, with a total $\mathrm{SFR}_\mathrm{UV+IR} = 267_{-153}^{+418}\,M_\odot\,\mathrm{yr}^{-1}$ that is almost completely obscured ($f_\mathrm{obs} = 0.94_{-0.09}^{+0.04}$). Using resolved ALMA observations in Bands 6, 8 and 9, we confirm both LBGs have hot dust temperatures ($T_d \approx 67 - 84\,\mathrm{K}$) and correspondingly high obscured fractions ($f_\mathrm{obs} \approx 0.88 -0.95$). We find the Western LBG to fall $\sim1\,\mathrm{dex}$ above the canonical IRX-$\beta_\mathrm{UV}$ relation, suggesting patchy dust obscuration. The compact Eastern component, on the other hand, is consistent with a Calzetti- or SMC-like dust screen within the uncertainties. Together with the similarly hot dust temperature recently reported for the $z=8.31$ galaxy MACS0416-Y1, our results suggest a non-negligible fraction of star formation at the bright end of the UV luminosity function is highly dust-obscured, even at $z\gtrsim7$.
\end{abstract}

\clearpage 




\section{Introduction}
\label{sec:introduction}

One of the key quantities describing the evolution of the galaxy population across time is the cosmic Star Formation Rate Density (SFRD). Our current census of the SFRD is heavily biased towards UV-luminous sources, i.e., towards star formation that is not obscured by dust \citep{madau2014,casey2018}. This bias becomes even more apparent beyond cosmic noon ($z\gtrsim4$), where the number density of infrared-luminous galaxies rapidly declines \citep[e.g.,][]{dudzeviciute2020,gruppioni2020,zavala2021}, and the obscured SFRD appears to mainly be driven by galaxies with lower typical dust contents \citep[e.g.,][]{khusanova2021,fujimoto2023}. Nevertheless, the contribution of these galaxies to the overall SFRD may be as high as $\sim30\%$ even at $z\sim7$ -- well into the Epoch of Reionization \citep[$z\gtrsim6$; e.g.,][]{fudamoto2021,algera2023,sun2025}.

While the number of dust-detected galaxies in the $z\gtrsim6$ Universe is steadily increasing -- primarily due to observations with the sensitive Atacama Large Millimeter/submillimeter Array \citep[ALMA, e.g.;][]{watson2015,hashimoto2019,tamura2019,faisst2020,bouwens2022,inami2022,schouws2022,witstok2022} -- most of these distant dusty galaxies remain detected in only a single ALMA band. Consequently, this leaves their dust properties, such as their dust mass ($M_d$), temperature ($T_d$) and emissivity index ($\beta_\mathrm{IR}$), woefully unconstrained. For most studies involving single-band detected galaxies at $z\sim7$, an ad hoc temperature of $T_d \sim 35-50$ K is therefore often used to obtain dust mass, infrared luminosity ($L_\mathrm{IR}$), and star formation rate (SFR) estimates \citep[e.g.,][]{schouws2022,algera2026}. Recent studies using two or multiple ALMA bands, however, have shown that the dust temperatures of galaxies may follow a wide distribution \citep[e.g.,][]{hashimoto2019,bakx2020,bakx2024,bakx2025,faisst2020,sugahara2021,algera2024,mitsuhashi2024}, suggesting such ad hoc assumptions could lead to strong biases in measured dust properties \citep[see e.g.,][]{bakx2021,algera2024b,lower2024,sommovigo_algera2025}. Indeed, obscured star formation rate measurements are highly sensitive to the dust temperature, since the IR luminosity varies as $\mathrm{SFR}_\mathrm{IR} \propto L_\mathrm{IR} \propto M_d T_d^{4+\beta_\mathrm{IR}}$, where typically $\beta_\mathrm{IR} \sim 2.0$ \citep[e.g.,][]{dacunha2021,bendo2025}. This opens up the possibility that the bright far-infrared emission detected by ALMA in some high-redshift galaxies may be attributed to warmer dust temperatures rather than unexpectedly large dust masses at this epoch \citep[][]{sommovigo2022b,choban2024}.

Accurately determining the presence of hot dust at $z\sim7$,\footnote{In this work, we refer to dust temperatures of $T_d \gtrsim 60\,\mathrm{K}$ as `hot', though we acknowledge that no standard definition exists.} however, is observationally challenging. First of all, this requires sampling the peak of the dust SED, necessitating the use of ALMA Bands 9 and/or 10 \citep[e.g.,][]{bakx2021}, which require notoriously good observing conditions ensuring a high phase stability. At the same time, it is important to simultaneously sample the Rayleigh-Jeans tail of the dust SED -- for example, through ALMA Bands 3 and/or 4 -- to alleviate degeneracies between the dust temperature and $\beta_\mathrm{IR}$ \citep[e.g.,][]{dacunha2021,algera2024b}. Such multi-band sampling is generally only feasible for the most infrared-luminous high-redshift systems with current facilities \citep[e.g.,][]{bakx2021,bakx2025,akins2022,witstok2023,algera2024b,algera2025_hz10}, limiting our understanding of dust properties in the early Universe.

Theoretical models suggest that galaxy dust temperatures increase towards earlier cosmic times, as a result of the typically powerful radiation fields and high star formation rate surface densities at these epochs \citep[e.g.,][]{liang2019,sommovigo2022,parente2026}. Observationally, such an increase in dust temperatures has indeed been inferred from stacking analyses out to $z\lesssim 4$ \citep[e.g.,][]{magnelli2014,schreiber2015}, although it remains unclear if these results can be directly extrapolated into the $z>6$ Universe. More recently, \citet{viero2022} performed a stacking analysis using photometrically-selected galaxies out to $z\sim9$ \citep[see also][]{casey2026}, suggesting a particularly rapid dust temperature evolution with a typical $T_d \sim 100\,\mathrm{K}$ by $z\approx8$. While the reliability of this result has been put into question \citep[e.g.,][]{sommovigo2022b}, and individual dust-detected galaxies at these epochs generally appear colder \citep[e.g.,][]{bakx2021,algera2024}, one high-redshift galaxy has now been shown to possess such a high dust temperature. This $z=8.31$ galaxy, known as MACS0416-Y1 (hereafter Y1), was first suggested to host hot dust ($T_d \gtrsim 85\,\mathrm{K}$) based on the dual-band ALMA observations presented in \citet{tamura2019} and \citet{bakx2020}. Further analysis in multiple ALMA bands -- including Band 9 -- has recently confirmed its temperature as $T_d = 91_{-35}^{+62}\,\mathrm{K}$ \citep[][]{bakx2025}. Recent work with the \textit{JWST} has suggested Y1 may be a merger \citep{harshan2024}, and possibly hosts a (weak) active galactic nucleus \citep[AGN; ][]{takechi2026}, although its warm dust emission is spatially extended \citep{tamura2023,bakx2025} and therefore likely heated primarily by star formation. At this stage, it remains unclear if the dust in Y1 is uniquely hot, or whether this galaxy may be representative of a population of hot, dusty galaxies in the Epoch of Reionization. Given the aforementioned strong dependence of the SFRD on the dust temperature, a hidden population of such hot galaxies could greatly boost the importance of obscured star formation in the $z\gtrsim6$ Universe \citep[e.g.,][]{viero2022,algera2023}.

The focus of this study is the UV and IR-luminous Lyman-break galaxy system B14-65666 at $z=7.15$ ($M_\mathrm{UV} = -22.4$), also known as the Big Three Dragons. The system was originally identified as a robust $z\sim7$ galaxy candidate \citep{bowler2012,bowler2014}, and was subsequently spatially resolved with the \textit{Hubble Space Telescope} by \citet{bowler2017}. These observations revealed a complex structure with two distinct components, indicating that the system may be a merger. Follow-up ALMA observations in Band 6 detected dust continuum emission from the Big Three Dragons \citep{bowler2018}, while the ALMA Band 6 and 8 observations presented in \citet{hashimoto2019} detected the \ciil{} and \oiiil{} lines from the galaxy after its redshift was spectroscopically confirmed based on the detection of Lyman-$\alpha$ \citep{furusawa2016}. The dual-band ALMA observations presented by \citet{hashimoto2019} moreover yielded robust dust continuum detections, and were expanded upon by \citet{sugahara2021}, who presented a new continuum detection in ALMA Band 7, as well as an upper limit on the \nii{} $122\,\mu\mathrm{m}$ line. Based on a modified blackbody fit to the three-point FIR SED, they inferred a largely unconstrained dust temperature of $T_d \approx 40 - 80\,$K, although based on energy-balance arguments, \citet{sugahara2021} suggested the temperature was likely on the upper end of this range. The Big Three Dragons have also been observed in ALMA Band 3 by \citet{hashimoto2023}, with the aim of detecting the CO(7-6) and CO(6-5) lines. However, no molecular gas nor the underlying continuum were detected.

In recent years, the Big Three Dragons have been extensively studied also with the \textit{James Webb Space Telescope} (\textit{JWST}). New \textit{JWST}/NIRCam observations presented in \citet{sugahara2024} confirmed the major-merger nature of the system, and studied the relation between its infrared excess ($\mathrm{IRX} = L_\mathrm{IR} / L_\mathrm{UV}$) and UV-continuum slope ($\beta_\mathrm{UV}$), the so-called IRX-$\beta_\mathrm{UV}$ relation \citep[see e.g.,][]{meurer1999,popping2017_irxbeta,bowler2024}. The high star formation rate of the system inferred by \citet{sugahara2024} -- and especially that of its Eastern component, which they find to be particularly compact -- suggests the Big Three Dragons is a powerful merger-induced starburst. As part of the GA-NIFS program, \citet{jones2024} presented \textit{JWST}/NIRSpec observations of the system, and inferred a metallicity of $Z\sim0.2-0.3\,Z_\odot$ for the global system. Recently, \citet{prieto-jimenez2025} presented new \textit{JWST}/MIRI medium-resolution spectroscopy of the galaxy pair which spatially resolved its H$\alpha$ emission. In agreement with the NIRCam-based analysis from \citet{sugahara2024}, they found different dust properties between the two components, suggesting the Eastern one has a high dust attenuation ($A_V \approx 1.5$), while the Western component appears nearly unobscured ($A_V \approx 0.1$).

In this work, we further explore the dust properties of the Big Three Dragons through new ALMA continuum observations. We simultaneously probe the peak and Rayleigh-Jeans tail of this suspected warm source, so as to accurately estimate its dust temperature, dust mass, and SFR. We present new data in ALMA Bands 4 and 9 and combine it with the existing ALMA observations of the Big Three Dragons reported in \citet{hashimoto2019, hashimoto2023} and \citet{sugahara2021}.

This paper is structured as follows. In Section \ref{sec:observations}, we present the data used in this work. The methods adopted for fitting the dust SED are explained in Section \ref{sec:methods}, and the results of this analysis are shown in Section \ref{sec:results}. Section \ref{sec:discussion} discusses the implications and is followed by our concluding remarks in Section \ref{sec:conclusions}. Throughout this work, we assume a standard $\Lambda$CDM cosmology, with $H_0=70\,\text{km\,s}^{-1}\text{\,Mpc}^{-1}$, $\Omega_m=0.30$ and $\Omega_\Lambda=0.70$. We further adopt a \citet{chabrier2003} IMF and determine obscured SFRs following \citet{inami2022} as $\mathrm{SFR}_\mathrm{IR}/(M_\odot\,\mathrm{yr}^{-1}) = 1.2 \times 10^{-10} L_\mathrm{IR}/L_\odot$.

\section{ALMA Data, calibration and imaging}\label{sec:observations}

We present new ALMA dust continuum observations of the Big Three Dragons in Bands 4 and 9, which we describe in detail below. These data complement previously taken observations of the galaxy in Bands 3 \citep{hashimoto2023}, 6 \citep{hashimoto2019}, 7 \citep{sugahara2021} and 8 \citep{hashimoto2019}, and we refer to these papers for a detailed description. Combined, the Band 3 through 9 observations cover a wavelength range of $\lambda_\mathrm{obs} \approx 0.45 - 3.3\,\mathrm{mm}$, which corresponds to $\lambda_\mathrm{rest} \approx 55-380\,\mu\mathrm{m}$ in the source rest-frame.

The new observations in Band 4 and 9 were carried out as part of the Cycle 10 ALMA program 2023.1.01033.S (PI Algera), which was designed to robustly measure the dust temperatures, masses and emissivity indices of five previously continuum-detected Lyman break galaxies through combined observations at the peak and Rayleigh-Jeans tail of the dust SED, following the strategy in \citet{algera2024b}. We here focus solely on the Big Three Dragons at $z=7.15$, being the most distant among the five targets. The ALMA Band 9 data of two additional $z\sim6$ galaxies taken as part of 2023.1.01033.S will be presented in Mitsuhashi et al.\ (in preparation), while the Band 4 observations targeting the $z=5.65$ galaxy HZ10 were previously combined with the Band 9 observations from \citet{villanueva2024} and Band 10 observations from 2024.1.01134.S (PI Herrera-Camus), as presented in \citet{algera2025_hz10}.

The Band 4 observations were carried out in five sessions between 2023 Dec 31 and 2024 Jan 01, for a total time of $4.9\,\mathrm{h}$ ($3.6\,\mathrm{h}$ on-source). The Band 9 observations were executed in two sessions on 2024 Nov 23 and 2024 Nov 25, and account for a total time of $136.4\,\mathrm{min}$, of which $63.5\,\mathrm{min}$ on-source.

\begin{figure*}
    \includegraphics[width=0.85\textwidth]{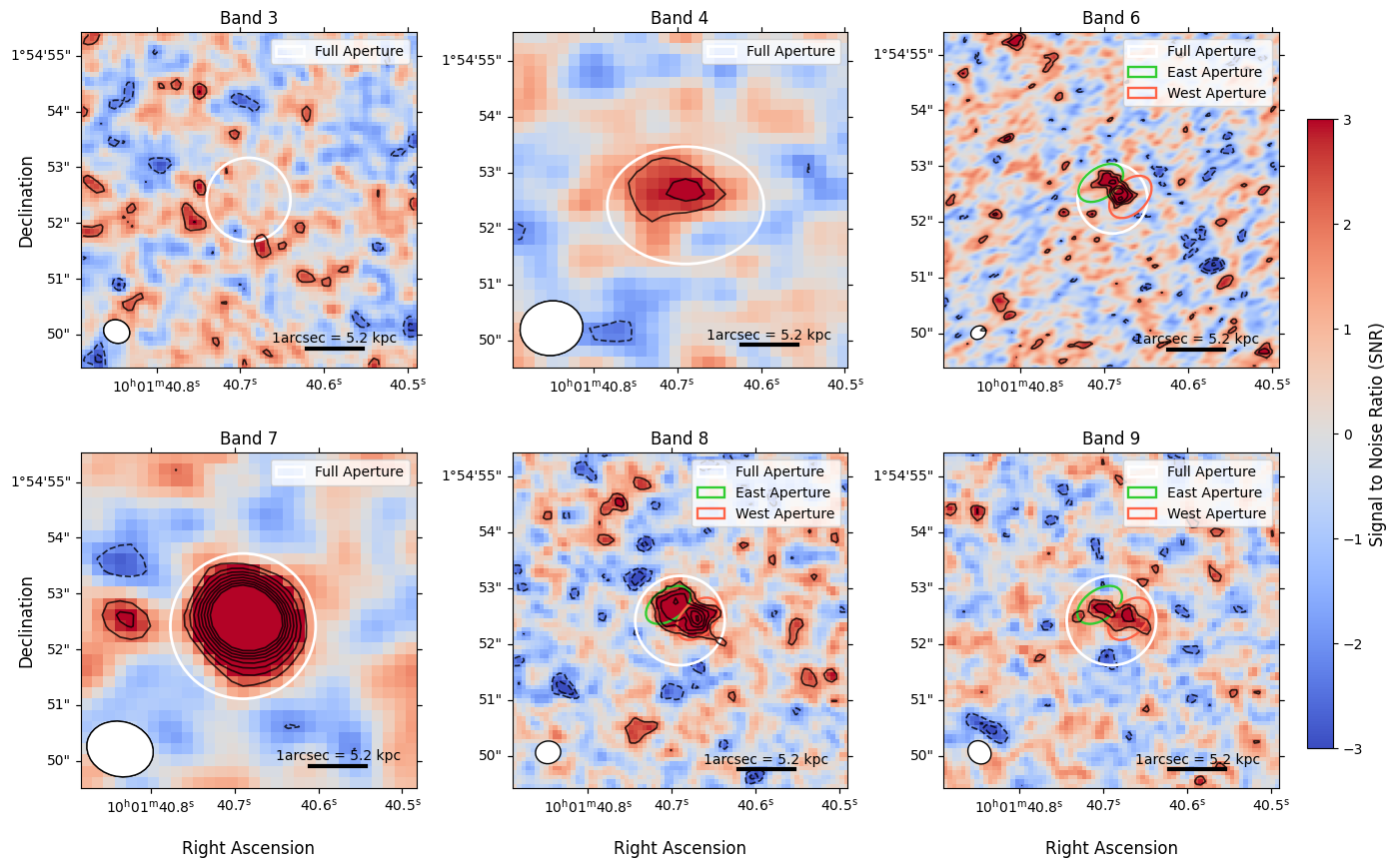}
    \caption{Cutouts ($6''\times 6''$) of the ALMA Band 3, 4 and 6 through 9 continuum observations of the Big Three Dragons. Solid contours (\textit{black}) are drawn from $2\sigma$ to $9\sigma$, increasing in steps of $1\sigma$. Dashed contours denote negative values, drawn from $-2\sigma$ downward in steps of $-1\sigma$. The Big Three Dragons remains undetected in Band 3, and is detected in all other bands at $\geq3\sigma$ significance. Aperture photometry was used to extract the fluxes of the full system (\textit{white }aperture) as well as the resolved East and West components (\textit{green }and\textit{ orange }apertures, respectively).} 
    \label{fig:cutouts}
\end{figure*}

\begin{figure*}
    \includegraphics[width=0.85\textwidth]{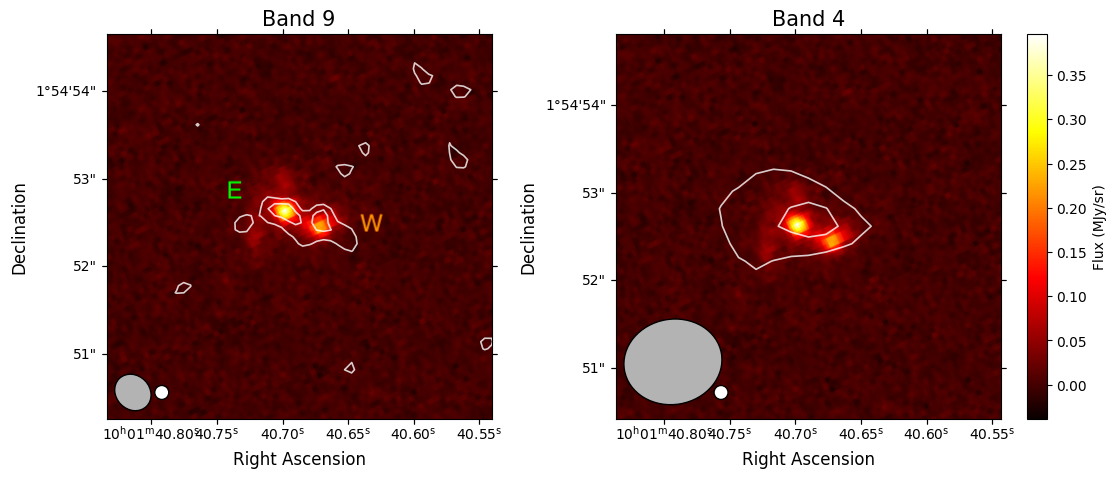}
    \caption{Band 9 (\textit{left}) and 4 (\textit{right}) contours, starting at $2\sigma$ and increasing in steps of $1\sigma$, are overlaid on the \textit{JWST}/NIRCam F444W image cutout ($\sim 4.4''\times 4.4''$)  of the Big Three Dragons from \citet{sugahara2024}. The \textit{JWST} data clearly resolve the Eastern and Western clumps, which align well with the Band 9 emission. The Band 4 emission is unresolved, though appears centered on the Eastern component. Overlays of Bands 6, 7, and 8 on NIRCam imaging can be found in \citet{sugahara2024}. The respective ALMA beams are shown in \textit{grey}, and the NIRCam F444W PSF is shown in \textit{white}.}
    \label{fig:jwst}
\end{figure*}

To ensure a consistent data reduction, we reduce the new and archival data in a homogeneous manner. All data are downloaded from the ALMA archive, and calibrated visibilities are restored following the standard approach using {\sc{scriptForPI.py}}. As the observations of the Big Three Dragons used in this work were carried out across different ALMA cycles -- spanning cycles 4 through 11 -- we ensure the version of {\sc{CASA}} \citep{casateam2022} used for the pipeline calibration and restoration of calibrated data are identical.

In producing the continuum images, we run the {\sc{tclean}} task on the calibrated visibility data in {\sc{CASA}} version 6.5.4. We ensure the channels containing the \cii{} and \oiii{} lines are excluded, removing those around the line center within $2\times$ the full-width at half-maximum (FWHM), which accounts for $98\%$ of the total line flux assuming a Gaussian profile. Neither the new Band 4 data -- covering the CO(10-9) and CO(9-8) lines -- nor the Band 3 observations presented in \citet{hashimoto2023} reveal any lines, and all channels are thus used in the imaging. Similarly, the Band 7 data cover but do not reveal the \nii{} $122\,\mu\mathrm{m}$ line \citep{sugahara2021}, and therefore all channels are included. Finally, the Band 9 observations did not target any emission lines. Briggs-weighted imaging is then carried out with a robust parameter of 2, providing a sensitivity and resolution akin to natural weighting. We will therefore henceforth refer to these as the `naturally-weighted images.' The beam sizes range from $\sim0.25$ to $\sim1.1\,$arcsec ($\sim1.3$ to $\sim5.7\,$kpc), with the Band 6 (Band 7) observations attaining the highest (coarsest) resolution.

We show the images in Figure \ref{fig:cutouts} and present their properties in Table \ref{tab:beam_params}. Dust continuum emission from the Big Three Dragons is clearly detected in the new Band 4 and 9 observations.
The peak-S/N in Band 4, in which the system remains unresolved, equals $3.5\sigma$. The Band 9 observations were conducted at $\sim0.4''$ resolution, and reveal two emission peaks at $3.7\sigma$ (Eastern) and $3.5\sigma$ significance (Western). We overplot the new ALMA observations on top of the \textit{JWST}/NIRCam F444W imaging from \citet{sugahara2024} in Figure \ref{fig:jwst}. We find that the ALMA peaks coincide with the positions of the two individual LBGs that together make up the Big Three Dragons. Both Band 9 peaks are cospatial with the \textit{JWST}/NIRCam counterparts within the $\sim 0.13"$ astrometric uncertainty, estimated as $\sim\mathrm{FWHM}/\mathrm{SNR}/0.9$. Given the spatially-resolved emission in the naturally-weighted Band 9 image, the peak S/N further increases upon $uv$-tapering the data, albeit at the expense of angular resolution. The nominal detection significance in a $0.8''$ tapered image, where the peak signal-to-noise is maximized, is $4.0 \sigma$ (Appendix \ref{app:taper}).

In agreement with previous works, dust emission is also detected in Bands 6, 7 and 8, while the continuum remains undetected in Band 3 \citep{hashimoto2018,hashimoto2023,sugahara2021}. The Big Three Dragons is moreover spatially resolved in ALMA Bands 6, 8, and 9, allowing us to perform a resolved analysis by fitting the 3-point SEDs of each clump in Section \ref{sec:results}. The clumps identified in Band 8 and Band 9 are cospatial at the current angular resolution of $\sim0.4''$, as may be expected since both bands trace warm dust. The Band 6 observations, on the other hand, show a slight spatial offset with respect to the higher-frequency data, peaking in the region between the two LBGs \citep{hashimoto2019,sugahara2024}. This offset may be because the Band 6 observations primarily trace colder and possibly more diffuse dust. To account for this, we define our photometric apertures in the following Section \ref{sec:methods} such that all emission is captured across bands.

\begin{table*}
\centering
\begin{tabular}{lcccccccc}
\hline
Band & Major & Minor & PA & RMS & $\lambda_\mathrm{obs}$ & Flux & Aperture Size & PID \\
\hline 
 & [arcsec] & [arcsec] & [deg] & [$\mu$Jy/beam] & [mm] & [$\mu$Jy] & [arcsec $\times$ arcsec] & \\
\hline
Band 3 & 0.475 & 0.411 & 66.2  & 4.5 & 3.26 & < $3 \times 12.8$ & 1.5 $\times$ 1.5 & 2018.1.01673.S \\
Band 4 & 1.124 & 0.975 & -80.2 & 5.2 & 2.23 & $23.8 \pm 10.6$ & 2.8 $\times$ 2.1 & 2023.1.01033.S \\
Band 6 & 0.284 & 0.230 & -60.9 & 8.7 & 1.32 & $136.9 \pm 33.3$ & 1.25 $\times$ 1.25 & 2016.1.00954.S \\
Band 7 & 1.200 & 0.981 & 75.2  & 9.8 & 1.02 & $211.6 \pm 19.6$ & 2.6 $\times$ 2.6 & 2019.1.01491.S \\
Band 8 & 0.452 & 0.405 & -80.0 & 27.4 & 0.73 & $467.1 \pm 85.4$ & 1.6 $\times$ 1.6 & 2016.1.00954.S, 2017.1.00190.S
 \\
Band 9 & 0.444 & 0.386 & 44.3  & 140.5 & 0.45 & $1162 \pm 452$ & 1.6 $\times$ 1.6 & 2023.1.01033.S \\
\hline
\end{tabular}

\caption{ALMA beam parameters and aperture flux measurements for each band. Note that the Big Three Dragons is not detected in Band 3, and an upper limit of $3\times$ the aperture flux uncertainty is used.}
\label{tab:beam_params}
\end{table*}

\begin{table}
    \centering
    \begin{tabular}{l|c|c}
    \hline
    Band & East Clump Flux & West Clump Flux \\
     & [$\mu$Jy] & [$\mu$Jy] \\
    \hline
    Band 6 & $63.0 \pm 19.2$ & $69.3 \pm 19.2$ \\
    Band 8 & $166.0 \pm 36.1$ & $191.0 \pm 36.1$ \\
    Band 9 & $381.1 \pm 191.3$ & $532.4 \pm 191.3$ \\
    \hline
    \end{tabular}
    \caption{Aperture flux measurements for the resolved east and west clumps in Bands 6,8 and 9.}
    \label{tab:clumpfluxes}
\end{table}

\section{Methods}
\label{sec:methods}

\subsection{Flux density measurements}

Fluxes and their corresponding uncertainties in each band were measured through aperture photometry using the Astropy \texttt{Photutils} package \citep{bradley2025}. Apertures were chosen through visual inspection while ensuring uniform central coordinates across all bands. The apertures for the global system, shown in white in Figure \ref{fig:cutouts}, are centered at the coordinates (10:01:40.69, +01:54:52.42) as reported by \citet{hashimoto2019}. The aperture coordinates for the east and west clumps are (10:01:40.70, +01:54:52.70) and (10:01:40.67, +01:54:52.45) respectively. For the east clump, an aperture of size $0.9" \times 0.54"$ with a position angle of $\mathrm{PA}=35^\circ$ was used, while for the west clump the aperture was $0.9" \times 0.58"$ with $\mathrm{PA}=45^\circ$. The band-wise aperture sizes for the global system as well as the corresponding flux density measurements for both the global system and the resolved clumps are listed in Table \ref{tab:beam_params} and Table \ref{tab:clumpfluxes}, respectively. Consistent with \citet{hashimoto2023}, no continuum emission is detected from the Big Three Dragons in ALMA Band 3, and we therefore adopt $3\times$ the error from the aperture flux density as an upper limit in the remainder of the paper. Both the global and component-wise fluxes in Bands 6, 8 and 9 were found to be fully consistent with the values reported in \citet{hashimoto2019} and \citet{sugahara2021}. As a sanity check, we also carry out a simultaneous 2D Gaussian fit for the two clumps with CASA \texttt{imfit}, and find the peak and integrated fluxes to be consistent with the component-wise fluxes derived from aperture photometry (comparison plots are provided in Appendix \ref{app:fluxes}). We account for flux calibration errors between the ALMA observations by adding a $5\%$ (Band 4), $10\%$ (Bands 6, 7, 8) and $20\%$ (Band 9) error in quadrature to the nominal uncertainties on the continuum fluxes, as recommended by the ALMA Technical Handbook \footnote{The technical handbook for Cycle 10 can be found at https://almascience.eso.org/documents-and-tools/cycle10}.

\subsection{Modified blackbody fitting}

Given that the Big Three Dragons are known to possess a moderate dust mass and relatively warm dust temperature, we assume the emission is optically thin \citep[following e.g.,][]{hashimoto2019,sugahara2021}. We note that assuming optically thick dust would yield a higher inferred dust temperature \citep[e.g.,][see also \citealt{kano2026} for a detailed discussion of optically thick dust in Y1]{faisst2020,algera2024b,lower2024}, and we return to this in Section \ref{sec:Tdust_Sigma}. With the current data, however, the wavelength where the dust turns optically thick cannot be constrained.

To fit the dust continuum SED, we adopt a single-temperature modified blackbody (MBB) function \citep[e.g.,][]{hildebrand1983,casey2012}: 

\begin{equation}
    S_{\nu} = \left(\frac{1+z}{d_{L}^2}\right)M_{d}\kappa_{0}\left(\frac{\nu}{\nu_{0}}\right)^{\beta_{\text{IR}}}[B_{\nu}(T_{d, z}) - B_{\nu}(T_{\text{CMB}, z})] \ .
\end{equation}

\noindent Here $d_L$ is the luminosity distance to redshift $z$, $M_d$ the dust mass, $\beta_\mathrm{IR}$ the dust emissivity index, $B_\nu$ the Planck function and $T_d$ the dust temperature. We adopt a dust opacity coefficient of $\kappa_0 = 10.41\,\mathrm{cm}^2\mathrm{g}^{-1}$ at $\nu_0 = 1900\,\mathrm{GHz}$ \citep[following e.g.,][]{sommovigo2022,algera2024}. We account for dust heating by the Cosmic Microwave Background (CMB) following \citet{dacunha2013}, and denote by $T_{d,z}$ the CMB-heated dust temperature, while $T_{d,0}$ is the equivalent temperature at a redshift $z=0$ (see equation 12 in \citealt{dacunha2013} to convert between the two). We also account for attenuation against the CMB via the term $B_\nu(T_{\mathrm{CMB},z})$, where $T_{\mathrm{CMB},z}$ is the CMB temperature at redshift $z$. At the redshift of the Big Three Dragons, $T_{\mathrm{CMB},z} = 2.73 \times (1 + z) \approx 22\,\mathrm{K}$.

We fit the MBB function using the \texttt{emcee} Python package \citep{foreman-mackey2013}, which implements the Markov Chain Monte Carlo (MCMC) method to explore the parameter space and determine the posterior distributions of the model parameters. The median (50th percentile) of the posterior distribution is then taken as the best-fit value, while the 16th and 84th percentiles define the lower and upper $1 \sigma$ uncertainties. For the dust mass and dust temperature, we adopt flat (uniform) priors: $ 4<\mathrm{log}(M_d/M_\odot)<12$ and $T_{\mathrm{CMB},z}<T_d<150$~K. We have verified that consistent results are obtained if we adopt a broader uniform prior. For the dust emissivity index $\beta_\mathrm{IR}$, we assume a Gaussian prior with a mean of 1.8 and a standard deviation of 0.5 \citep[following e.g.,][]{algera2024,mitsuhashi2024}. However, we also explore an MBB fit with a fixed $\beta_\mathrm{IR} = 2.0$ in Appendix \ref{app:fixed_beta2}. Upper limits are included in the fitting following \citet[][see also \citealt{bakx2020,algera2024}]{sawicki2012}.

We find in Section \ref{sec:results} that the Band 9 observations do not yet probe the peak of the dust SED. To better constrain the MBB fit, we therefore use \textit{Herschel}/PACS and SPIRE upper limits at $100, 160, 250$ and $350\,\mu\mathrm{m}$, having confirmed the galaxy is not detected in \textit{Herschel} imaging.\footnote{We do not use the \textit{Herschel}/SPIRE upper limit at $500\,\mu\mathrm{m}$ as a similar wavelength is covered by our (significantly deeper) Band 9 observations.} We adopt $3\times$ the typical noise in the \textit{Herschel} COSMOS maps obtained by \citet{jin2018} for the upper limits. Finally, we note that the infrared luminosity $L_\mathrm{IR}$ is determined by integrating MBBs drawn from a representative subset of the fitting posteriors across $8-1000\,\mu\mathrm{m}$. 

\section{Results}
\label{sec:results}

\begin{figure*}
	\begin{center}
		\includegraphics[width=0.9\textwidth]{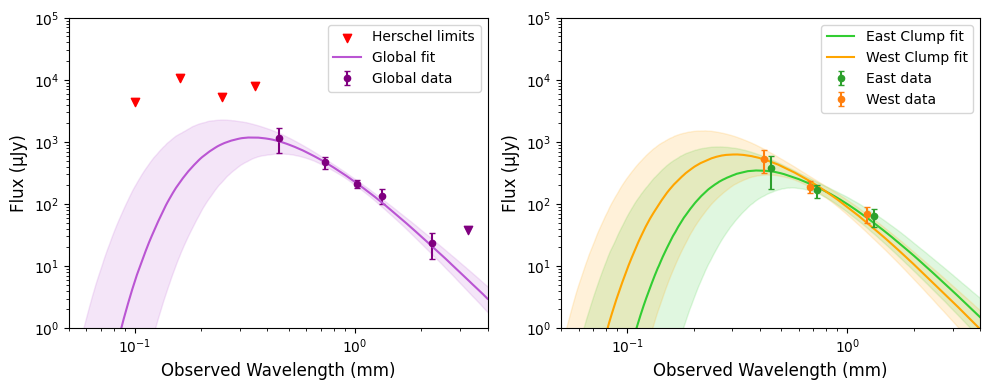}
		\caption{A comparison between the global and component-wise MBB fits. The \textit{left} panel contains the global  MBB fit (\textit{purple}) to the multi band ALMA data, with the Band 3 (\textit{purple triangle})  and Herschel data points (\textit{red triangles}) used as upper limits. The solid line represents the median fit while the shaded region covers the 16th-84th percentiles.The \textit{right} panel shows the MBB fits to the resolved data from ALMA Bands 6, 8 and 9 for the Eastern (\textit{green}) and Western (\textit{orange}) components, respectively. The data points for the Western component have been shifted to the left by a factor of 0.93 for clarity.} 
        \label{sed}
	\end{center}
\end{figure*}

We show a fit to the global dust SED of the Big Three Dragons, covered in Bands 3, 4 and 6 through 9, in the left panel of Figure \ref{sed}. We moreover show a corner plot highlighting the fitting posteriors and parameter degeneracies in Figure \ref{fig:corner} in Appendix \ref{app:corner}.

The far-IR flux density of the Big Three Dragons continues to rise towards shorter wavelengths, with the Band 9 data probing $\lambda_\mathrm{rest} \approx 55\,\mu\mathrm{m}$ seemingly still not completely capturing the peak of the SED. This is suggestive of hot dust, which is confirmed by the MBB fit yielding a high dust temperature of $T_d = 78^{+35}_{-23}\,$K. We moreover infer a total dust mass of $\log(M_{\text{d}}/M_\odot) = 6.85^{+0.33}_{-0.27}$, and an emissivity index of $\beta_\mathrm{IR} = 1.47^{+0.46}_{-0.35}$. The latter, however, is difficult to tightly constrain with the present data, given the modest S/N of the Band 4 detection. For that reason, we also perform a fit with a fixed $\beta_\mathrm{IR} = 2.0$ in Appendix \ref{app:fixed_beta2}. As shown in Figure \ref{fig:beta2}, this yields a lower nominal dust temperature of $T_d(\beta_\mathrm{IR}=2.0) = 57_{-13}^{+15}\,\mathrm{K}$, although this remains consistent with our fiducial result within the uncertainties. Similarly, the inferred dust mass and IR luminosity from this fixed-$\beta_\mathrm{IR}$ fit remain consistent with a fit where $\beta_\mathrm{IR}$ is left free (Table \ref{tab:sed_2.0}). Given that the current Band 4 data are already quite deep, the most efficient way of more accurately constraining the emissivity index of the Big Three Dragons in the future is through additional observations in ALMA Band 5, where the galaxy is considerably brighter. For the time being, we proceed with our fiducial fit where $\beta_\mathrm{IR}$ is left free, and compare our measurements to previous studies of the Big Three Dragons, as well as to other high-redshift galaxies observed with ALMA, in Section \ref{sec:discussion}. 

To assess the impact of the \textit{Herschel} upper limits on the inferred dust properties, we repeated the modified blackbody fit excluding the \textit{Herschel} constraints. Without the upper limits, the posterior shifts toward a slightly higher dust temperature as expected, with broader uncertainties extending to higher temperatures ($T_d = 86_{-29}^{+50}\,\mathrm{K}$) and correspondingly higher values for $L_{\mathrm{IR}}$ and SFR$_\mathrm{IR}$. These differences are modest and remain fully consistent with our fiducial fit within the uncertainties, indicating that the \textit{Herschel} non-detections primarily act to rule out the warmest dust-temperature solutions.

As discussed in Section \ref{sec:observations}, the observations in ALMA Bands 6, 8 and 9 resolve the two main components of the Big Three Dragons, and their flux densities are extracted in the two elliptical apertures overplotted in Figure \ref{fig:cutouts}. We show two MBB fits to these components in the right panel of Figure \ref{sed}. We infer that the Eastern clump has a dust temperature of $T_d = 67^{+44}_{-26}$ K and a dust mass of $\log(M_{\text{d}}/M_\odot) = 6.64^{+0.64}_{-0.47}$, with an emissivity index of $\beta_\mathrm{IR} = 1.48^{+0.52}_{-0.47}$. The Western clump appears to exhibit a slightly higher dust temperature of $T_d = 84^{+40}_{-30}$ K, as a result of its higher Band 9 flux density compared to the Eastern clump. However, due to the large uncertainties, we cannot conclusively establish which component is warmer. The Western clump contains a dust mass of $\log(M_\mathrm{d}/M_\odot) = 6.37^{+0.46}_{-0.33}$ and has a similar emissivity index of $\beta_\mathrm{IR} = 1.65^{+0.47}_{-0.45}$. Given the current available data, we thus conclude that both clumps have similar dust emission properties.

\begin{table*}
\centering
\begin{tabular}{lccccccccc}
\hline
Component &
$\log(M_{\text{d}}/M_\odot)$ &
Temperature &
$\beta_\mathrm{IR}$ &
$\log(L_{\mathrm{IR}}/L_\odot)$ &
SFR$_\mathrm{IR}$ &
$L_{\mathrm{UV}}$* &
SFR$_\mathrm{UV}$** &
$f_{\mathrm{obs}}$ &
$\log$ IRX \\
&
&
[K] &
&
&
[$M_\odot\,\mathrm{yr}^{-1}$] &
[$10^{21}\,\mathrm{W\,Hz^{-1}}$] &
[$M_\odot\,\mathrm{yr}^{-1}$] &
&
\\
\hline
Full System & $6.85^{+0.33}_{-0.27}$ & $78^{+35}_{-23}$ & $1.47^{+0.46}_{-0.35}$ & $12.32^{+0.43}_{-0.41}$ & $251^{+418}_{-153}$ & $22.6\pm1.1$ & $16.0\pm0.8$ & $0.94^{+0.04}_{-0.09}$ & $1.25^{+0.46}_{-0.28}$ \\
[1ex]
East Clump  & $6.64^{+0.64}_{-0.47}$ & $67^{+44}_{-26}$ & $1.48^{+0.52}_{-0.47}$ & $11.71^{+0.58}_{-0.49}$ & $61^{+171}_{-41}$ & $11.8\pm0.7$ & $6.9\pm0.4$ & $0.88^{+0.07}_{-0.30}$ & $0.92^{+0.55}_{-0.35}$ \\
[1ex]
West Clump  & $6.37^{+0.46}_{-0.33}$ & $84^{+40}_{-30}$ & $1.65^{+0.47}_{-0.45}$ & $12.07^{+0.53}_{-0.51}$ & $141^{+340}_{-98}$ & $9.7\pm0.6$ & $8.4\pm0.5$ & $0.95^{+0.03}_{-0.11}$ & $1.37^{+0.54}_{-0.34}$ \\
[1ex]
\hline
\end{tabular}

\caption{Physical properties of the full system and individual clumps derived from MBB fitting. (*) UV luminosity at 1500\AA \ from \citet{prieto-jimenez2025} (**) Derived following \citet{inami2022} as SFR$_\mathrm{UV} /(M_\odot \ \mathrm{yr}^{-1}) = 7.1 \times 10^{-29} L_{\nu}/(\mathrm{erg \ s^{-1}Hz^{-1}})$.}
\label{tab:sed_results}
\end{table*}

\section{Discussion}
\label{sec:discussion}

\subsection{Hot dust in the Big Three Dragons}
\label{sec:discussionTemperature}

\subsubsection{A comparison to previous studies of the system}

We start by comparing the global dust temperature we infer for the Big Three Dragons of $T_d \approx 78\,\mathrm{K}$ to previous studies of the system. \citet{hashimoto2019} first presented ALMA Band 6 and 8 observations of the Big Three Dragons, and inferred $T_d = 48 - 61\,\mathrm{K}$ based on MBB-fitting with a fixed $\beta_\mathrm{IR} = 1.5, 2.0$. A follow-up study by \citet{sugahara2021}, incorporating a new Band 7 continuum detection, yielded a nominal temperature of $T_d = 41 - 54\,\mathrm{K}$ given the same set of $\beta_\mathrm{IR}$, though with uncertainties spanning a wide range of $T_d = 29 - 96\,\mathrm{K}$. They moreover inferred the dust temperature of the Big Three Dragons using the radiative equilibrium models from \citet{inoue2020}. Assuming either a spherical shell or homogeneous sphere geometry, \citet{sugahara2021} suggested the dust temperature was likely $T_d = 80 - 100\,\mathrm{K}$, although adopting a clumpy geometry would reduce this to a more modest $T_d = 40 - 80\,\mathrm{K}$. Making use of full UV-to-FIR SED-fitting including new \textit{JWST}/NIRCam observations, \citet{sugahara2024} modeled the two clumps of the Big Three Dragons separately, and inferred a possible high dust temperature for the Eastern component of $T_d \gtrsim 65\,\mathrm{K}$, although they suggested the Western component to host colder dust based on its offset from the IRX-$\beta_\mathrm{UV}$ relation, which we discuss in further detail in Section \ref{sec:discussionObscuredSF}. Altogether, previous ALMA observations thus hinted at, but could not definitively establish, the presence of hot dust in the Big Three Dragons. Our new Band 9 and 4 observations now strongly indicate a high global dust temperature in the range of $55 \mathrm{K} - 112\mathrm{K}$, with a median estimate of $78 \mathrm{K}$.

\begin{figure}
	\begin{center}
		\includegraphics[width=0.95\columnwidth]{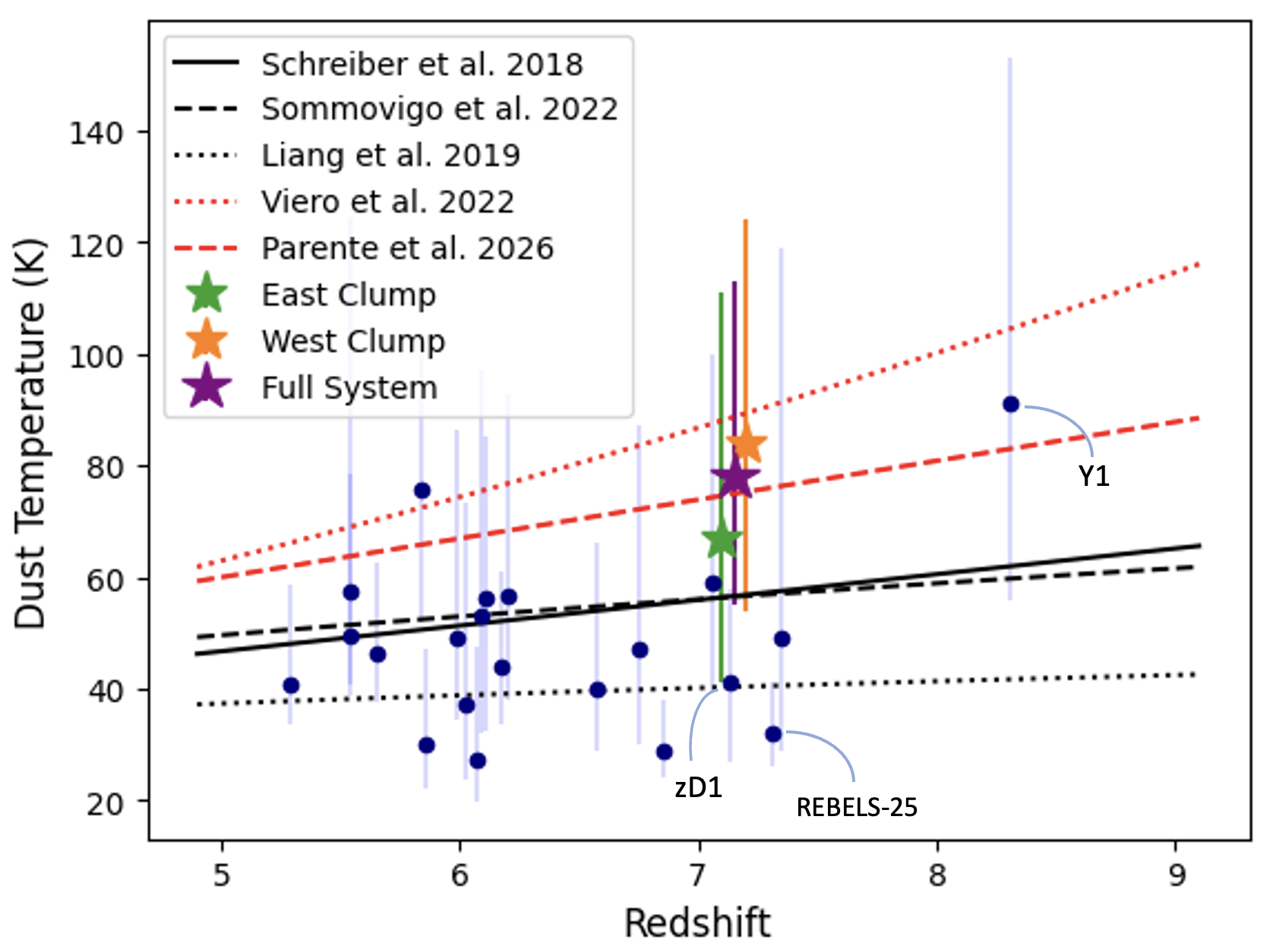}
		\caption{The cosmic evolution of dust temperatures based on high-redshift Lyman-break galaxies with dual- and multi-band ALMA detections \citep[\textit{blue} points;][]{faisst2020,bakx2021,bakx2025,akins2022,witstok2022,algera2024,algera2024b,algera2025_hz10,mitsuhashi2024}. The measured $T_{d}$ of the Big Three Dragons system (\textit{purple star}) as well as the east-west clumps (\textit{green} and \textit{orange star}, respectively; slightly offset in redshift for clarity) lie well above the general $z\gtrsim5.5$ population. Theoretical and empirical relations between $T_{d}$ and $z$ (\textit{dotted, dashed} and \textit{solid lines} -- references are provided in the legend) are plotted alongside for comparison. For ease of comparison, we also explicitly label all known galaxies at $z \gtrsim 7$ with Band 9 flux measurements: REBELS-25, A1689-zD1 and MACS0416-Y1. The Big Three Dragons is among the hottest known galaxies at this epoch.} 
        \label{fig:tdust}
	\end{center}
\end{figure}

\subsubsection{The dust temperature $-$ redshift relation}

We proceed by placing our constraints on the dust SED of the Big Three Dragons into the context of other high-redshift Lyman-break galaxies (LBGs) with dual- and multi-band dust continuum measurements. We focus first on the global SED of the system, while discussing its two individual components in more detail below. The dual- and multi-band dust temperature measurements are drawn from a variety of studies in the literature \citep{bakx2020,bakx2021,bakx2025,faisst2020,akins2022,witstok2022,algera2024,algera2024b,algera2025_hz10,mitsuhashi2024,villanueva2024}, and are plotted as a function of redshift in Figure \ref{fig:tdust}.

Most of the LBGs in our comparison sample, which span $z\approx5.4-8.3$, have dust temperatures of $T_d \approx 40 - 50\,\mathrm{K}$, well below the temperature inferred for the Big Three Dragons. Among this sample, two galaxies at a similar redshift as our target -- A1689-zD1 at $z=7.13$ and REBELS-25 at $z=7.31$ -- also benefit from well-constrained dust SEDs up to ALMA Band 9, though have significantly colder temperatures of $T_d \approx 32 - 45\,\mathrm{K}$ \citep{bakx2021,algera2024b}. Beyond our target, the only other $z > 7$ galaxy with a similarly well-constrained dust SED that suggests a hot dust temperature is the $z=8.31$ galaxy MACS0416-Y1. This galaxy -- which remains the most distant dust detection to date -- was initially detected in ALMA \oiiil{} and underlying continuum by \citet{tamura2019}, with follow-up observations from \citet{bakx2020} providing a \ciil{} detection and an upper limit on the underlying continuum flux density. Recently, \citet{bakx2025} detected this galaxy in ALMA Band 9, as well as in deeper Band 5 observations, confirming its hot dust temperature of $T_d = 91_{-35}^{+62}\,\mathrm{K}$. This is similar to the value inferred for the Big Three Dragons, although our target is more luminous -- a point that we will return to in Section \ref{sec:discussionObscuredSF}.

We proceed by comparing the observed dust temperature for the Big Three Dragons to empirical relations and theoretical models. Several functions describing the relationship between dust temperature and redshift have been proposed in the literature \citep[e.g.,][]{schreiber2018,liang2019,sommovigo2022,viero2022, parente2026}, although their predictions vary widely for redshifts $z>6$.\footnote{We note that these works predict/measure the `effective' dust temperature, and can thus readily be compared to our measurements inferred through MBB fitting. This effective temperature corresponds to neither the luminosity- nor the mass-weighted temperature, which can be regarded as `physical' temperatures of the dust. We refer to \citet{liang2019} and \citet{sommovigo_algera2025} for a detailed discussion.} The large uncertainties on our dust temperature measurement make it consistent with most published $T_d-z$ relations. However, at face value, the median temperature agrees best with the \citet{parente2026} relation, bounded on the lower side by the theoretical predictions of \citet{sommovigo2022}, while the upper uncertainty exceeds even the relatively steep empirical relation reported by \citet{viero2022}.

While samples remain small, the fact that now two $z>7$ galaxies have robust multi-band measurements revealing a hot dust temperature suggests that this could be a rare but important phase in early galaxy evolution.

\begin{figure}
	\begin{center}
		\includegraphics[width=0.95\columnwidth]{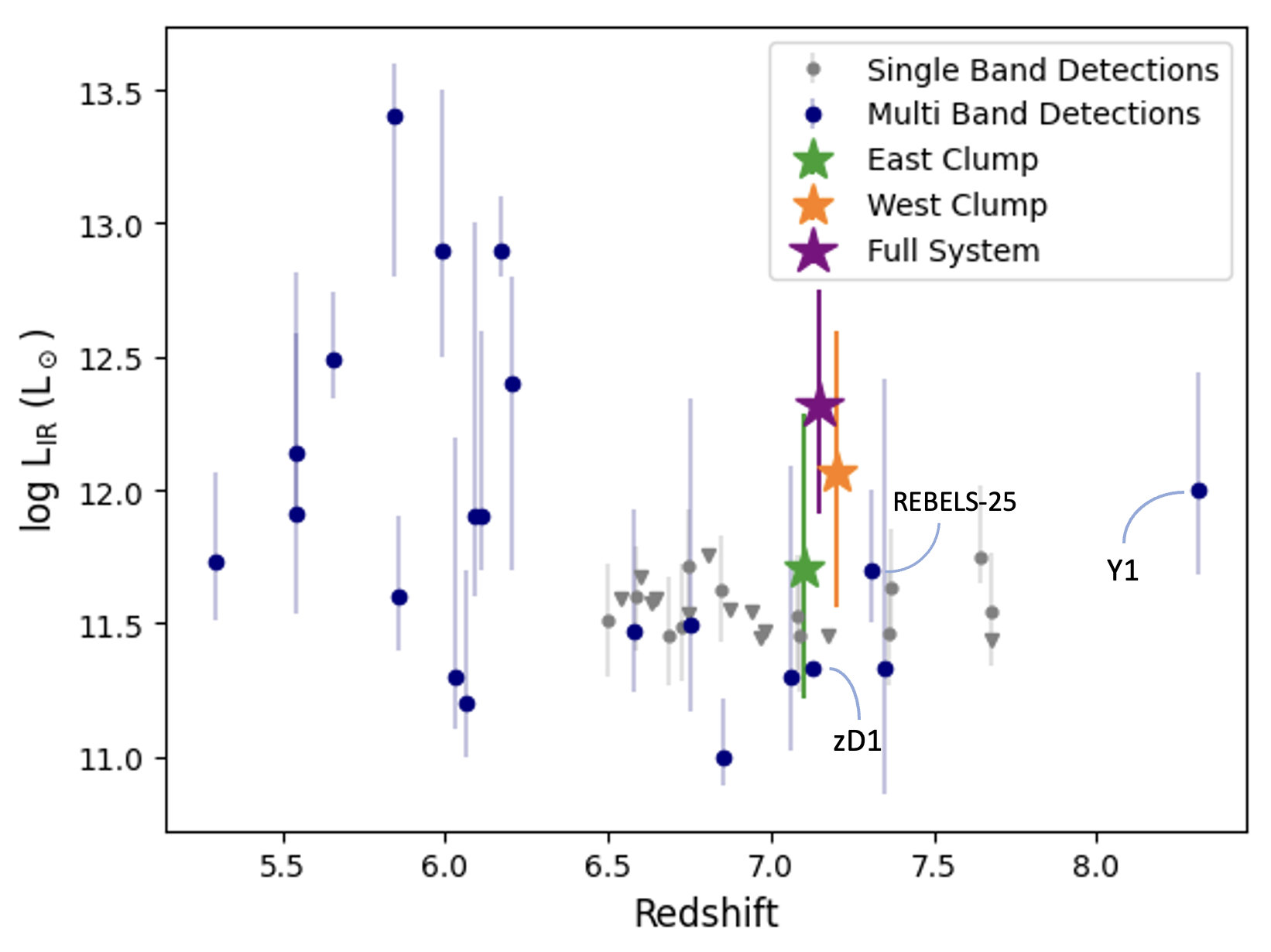}
		\caption{The IR luminosity $L_{\mathrm{IR}}$ of $z\gtrsim5$ Lyman-break galaxies targeted with ALMA as a function of redshift. The \textit{purple, green} and \textit{orange stars}, respectively, represent the $L_\mathrm{IR}$ of the Big Three Dragons system as well as its east-west clumps. The luminosities of literature galaxies detected and constrained across multiple bands are shown in \textit{blue} (see Fig \ref{fig:tdust} for references). Single-band detected REBELS galaxies \citep[][]{bouwens2022, inami2022}, whose luminosities were derived from an assumed fixed $T_d \approx 48\,\mathrm{K}$, are shown in \textit{grey}, with downward triangles representing $3\sigma$ upper limits. The Big Three Dragons stands out as a uniquely IR-luminous Lyman break galaxy at $z\gtrsim7$.} 
        \label{fig:lir_redshift}
	\end{center}
\end{figure}

\subsection{Obscured star formation}
\label{sec:discussionObscuredSF}

\subsubsection{The obscured fraction $-$ stellar mass relation}

The total infrared luminosity ($L_{\mathrm{IR}}$) of the Big Three Dragons, integrated across $8 - 1000 \mu$m, is found to be $\log(L_\mathrm{IR}/L_\odot) = 12.32_{-0.41}^{+0.43}$ for the global system, which is high as expected given its hot dust temperature. When plotted against other high-redshift LBGs, as shown in Figure \ref{fig:lir_redshift}, our target stands out as the most IR-luminous LBG observed at this epoch. The western clump, in particular, is classified as an Ultra-luminous Infrared Galaxy (ULIRG), given that its infrared luminosity exceeds $L_\mathrm{IR} > 10^{12}\,L_\odot$, while the slightly less luminous Eastern component falls into the LIRG regime (Table \ref{tab:sed_results}). In this literature comparison, we include single-band–detected galaxies from the REBELS survey \citep{bouwens2022,inami2022}, whose $L_{\mathrm{IR}}$ values are derived assuming typical dust temperatures of $T_d\approx48\,\mathrm{K}$, as well as the dual/multi-band-detected sources introduced in Section \ref{sec:discussionTemperature}. At $z\gtrsim6$, the only LBGs with a comparable IR luminosity to the Big Three Dragons can be found in the UV-luminous sample from \citet{mitsuhashi2024}, although these are found at a lower redshift of $z\approx6$ and are only detected in two ALMA bands. At slightly lower redshift, the $z=5.65$ LBG HZ10, whose IR luminosity is constrained through observations spanning Bands 4 through 10, is also similarly IR-luminous as our target \citep[][]{villanueva2024,algera2025_hz10}. While the $z=8.31$ galaxy Y1 has a similar dust temperature as the Big Three Dragons, it is overall less luminous with $\log(L_\mathrm{IR}/L_\odot) \approx 12.0$. However, we note that the uncertainties on the individual IR luminosities of high-redshift LBGs remain large, as even with high-frequency observations, accurately constraining the peak of the IR emission and thus $L_\mathrm{IR}$ is difficult in the case of hot dust \citep[e.g.,][]{sommovigo_algera2025}.

Although the Big Three Dragons appears more IR-luminous than any currently known $z>7$ Lyman-break galaxy, we note that quasars at this epoch can show a comparable or even higher $L_\mathrm{IR}$ \citep[e.g.,][]{fujimoto2022,tripodi2024,salvestrini2025}. However, the extent to which their dust is heated by star formation in the host galaxy or by radiation from the central black hole remains debated \citep[e.g.,][]{walter2022,meyer2025,silverman2025,tadaki2025}. Our target does not exhibit any clear sign of AGN activity in \textit{JWST} observations \citep{sugahara2024,jones2024,prieto-jimenez2025}, suggesting its infrared brightness may be from a purely star-forming origin (however, see Section \ref{sec:Tdust_Sigma} where we discuss whether the Eastern component could host an obscured AGN). Still, this is not completely unprecedented. At a slightly lower redshift, the bright infrared-selected starburst SPT0311-58 at $z=6.900$ similarly does not show any clear sign of AGN activity \citep{alvarez-marquez2023,arribas2024}, but has an even higher IR-luminosity well in excess of $\log(L_\mathrm{IR}/L_\odot) > 13$ \citep{strandet2017,marrone2018}. Unlike the Big Three Dragons, however, this source has been discovered in a wide $2500\,\mathrm{deg}^2$ area \citep{vieira2013,reuter2020}, while our target was discovered in the much smaller $\sim2\,\mathrm{deg}^2$ COSMOS field \citep{bowler2014}.

\begin{figure}
	\begin{center}
		\includegraphics[width=0.95\columnwidth]{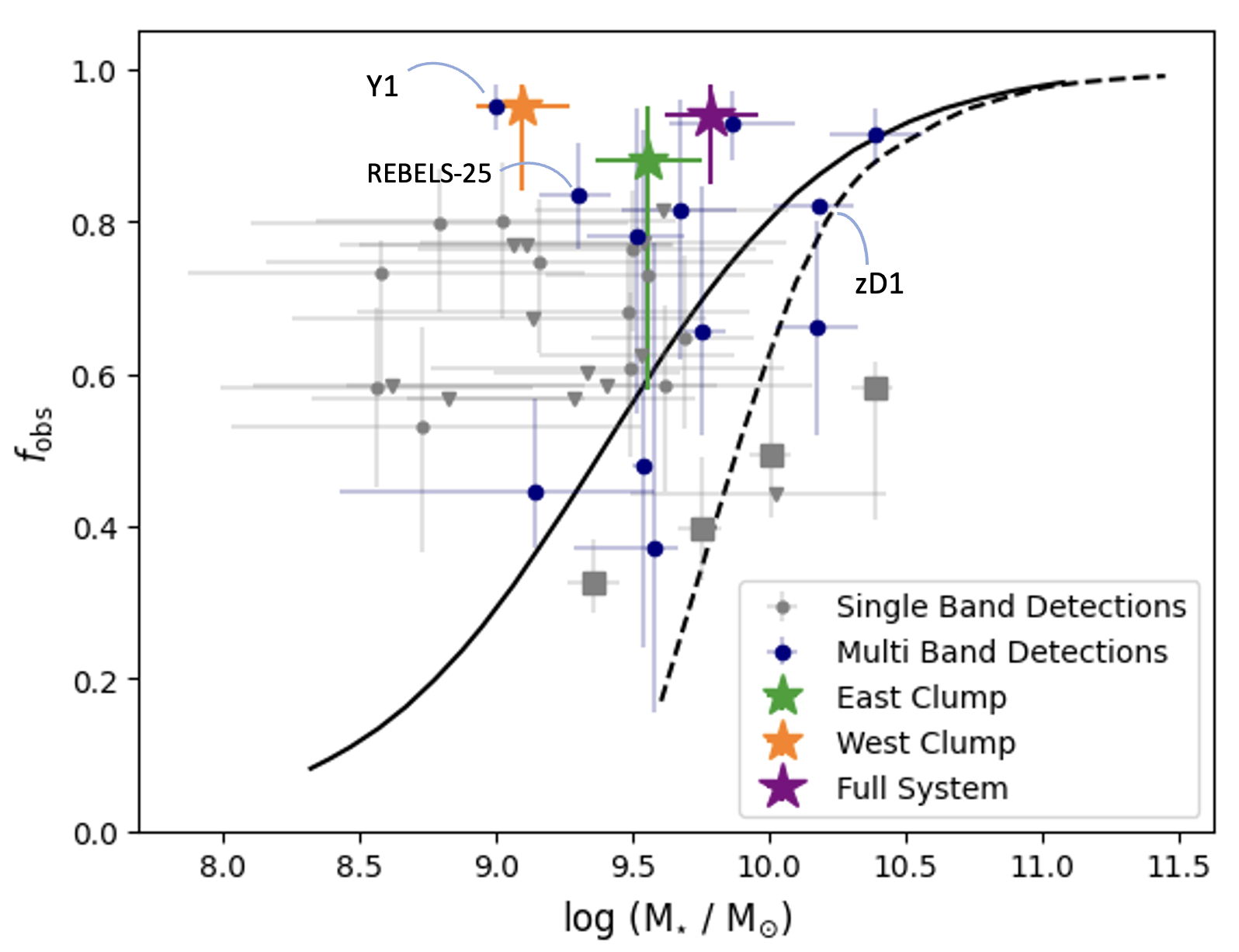}
		\caption{Obscured fraction versus the stellar mass of the Big Three Dragons compared against multi-band and single-band detected galaxies at high redshifts ($z\gtrsim 5.5$; see Fig \ref{fig:tdust} and \ref{fig:lir_redshift} for references -- only sources with $M_\star$ measurements in the literature are plotted). The \textit{solid} and \textit{dashed} lines \citep{whitaker2017} represent the trend observed in $z = 0 - 2.5$ galaxies assuming two different far-infrared emission templates. The grey squares represent single-band stacks from \citet{algera2023} with an assumed temperature of $T_d \approx 45\,\mathrm{K}$. The Big Three Dragons, as well as its two individual components, fall well above the low-redshift relation.} 
        \label{fig:fobs}
	\end{center}
\end{figure}

From the infrared luminosity of the Big Three Dragons, we derive an obscured star formation rate of $\mathrm{SFR}_\mathrm{IR} = 251_{-153}^{+418}\,M_\odot\,\mathrm{yr}^{-1}$ using the relation in Section \ref{sec:introduction}. Combined with its UV-based $\mathrm{SFR}_\mathrm{UV} \approx 16\,M_\odot\,\mathrm{yr}^{-1}$ \citep[][see also Table \ref{tab:sed_results}]{prieto-jimenez2025}, we thus infer a total star formation rate of $\mathrm{SFR}_\mathrm{UV+IR} = 267_{-153}^{+418}\,M_\odot\,\mathrm{yr}^{-1}$. This is slightly higher than, although fully consistent with, the star formation rate previously inferred by \citet{sugahara2024} from UV to far-IR SED fitting ($\mathrm{SFR}_{10\,\mathrm{Myr}} = 207_{-51}^{+65}\,M_\odot\,\mathrm{yr}^{-1}$), although about twice as high as the SFR inferred by \citet{prieto-jimenez2025}. As discussed in detail in those two works, the higher SFR inferred by \citet{sugahara2024} is likely primarily due to the larger adopted photometric apertures compared to \citet{prieto-jimenez2025}, which captures more of the total emission from the system.

From the combined IR and UV star formation rates, we can estimate the obscured fraction, $f_{\mathrm{obs}} = \mathrm{SFR}_\mathrm{IR} / \mathrm{SFR}_\mathrm{UV + IR}$ that quantifies the fraction of total star formation in a galaxy that is hidden by dust. At low redshift, $f_\mathrm{obs}$ is known to be a strongly increasing function of stellar mass \citep[e.g.,][]{whitaker2017}. The dustiest galaxies among the ALPINE and REBELS surveys also follow the $f_\mathrm{obs} - M_\star$ relation \citep[][]{fudamoto2020,inami2022,fisher2026}, although stacking analyses including also the dust-undetected sources targeted in these surveys reveal lower average obscured fractions \citep{fudamoto2020,algera2023}. On the other hand, there is some evidence that $z\gtrsim6$ galaxies not pre-selected in the UV show a much higher $f_\mathrm{obs}$, suggesting a population of particularly obscured sources still exists at these epochs \citep[e.g.,][]{fudamoto2021,endsley2023,bakx2024,vanleeuwen2024}. 

The global system has an obscured fraction of $f_\mathrm{obs} = 0.94^{+0.04}_{-0.09}$ (Figure \ref{fig:fobs}), which places the Big Three Dragons well above the $f_\mathrm{obs} - M_\star$ relation from \citet{whitaker2017} established for $z < 2.5$ galaxies. The two individual components, whose masses were obtained from resolved SED-fitting in \citet{sugahara2024}, similarly lie above the \citet{whitaker2017} relation, with the Eastern and Western ones showing an obscured fraction of $f_\mathrm{obs} = 0.88^{+0.07}_{-0.30}$ and $0.95^{+0.03}_{-0.11}$, respectively. The value inferred for the Western clump is particularly high given its modest stellar mass of $\log(M_\star/M_\odot) \approx 9.09$ \citep{sugahara2024}. \\

\begin{figure}
	\begin{center}
		\includegraphics[width=0.95\columnwidth]{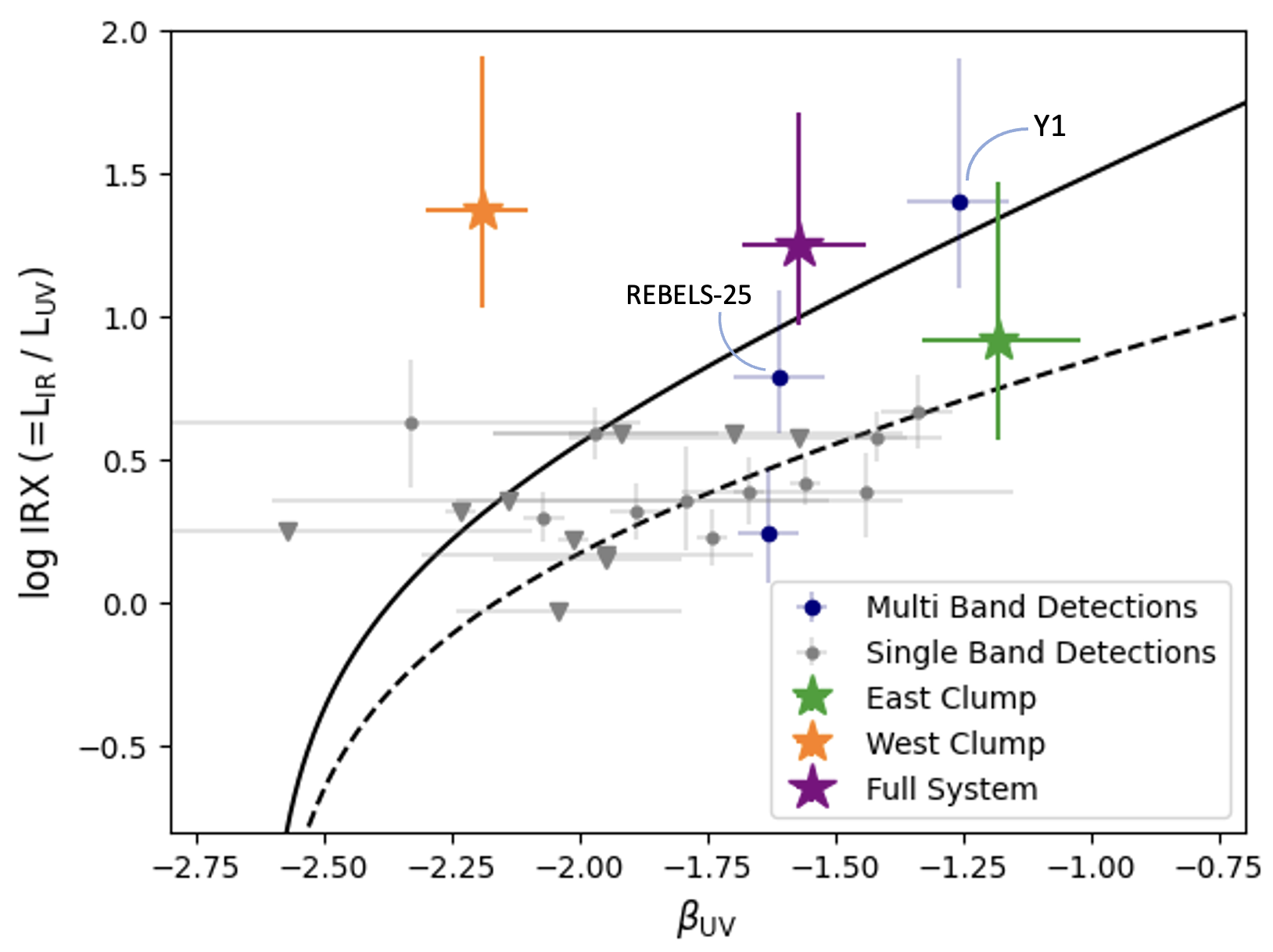}
		\caption{The Big Three Dragons (\textit{purple star}) and its two components (\textit{orange} and \textit{green stars}) on the IRX-$\beta_\mathrm{UV}$ relation. The solid (dashed) line represents the Calzetti (SMC) relation anchored to a $\beta_\mathrm{UV,0} = -2.62$. The \textit{grey} points are $z\approx7$ galaxies from the REBELS survey \citep{bowler2024}, for which a fixed dust temperature of 46 K is assumed. \textit{Blue} points correspond to multi-band measurements of $z > 6.5$ galaxies from \citet{algera2024,algera2024b} and \citet{bakx2025}. The Eastern component, which shows a reddened $\beta_\mathrm{UV}$, falls in between the canonical Calzetti and SMC relations, while the Western clump lies $\gtrsim1\,\mathrm{dex}$ above either.} 
        \label{fig:irx_beta}
	\end{center}
\end{figure}

\subsubsection{The $IRX-\beta_\mathrm{UV}$ relation}

The effects of dust obscuration can be quantified both by how much it attenuates and reddens the rest-frame UV and optical emission of a galaxy, and by the overall output in the far-infrared. This forms the basis of the $\mathrm{IRX}-\beta_\mathrm{UV}$ relation, which compares the infrared excess $\mathrm{IRX} = L_\mathrm{IR} / L_\mathrm{UV}$ and the UV continuum slope $\beta_\mathrm{UV}$. The $\mathrm{IRX}-\beta_\mathrm{UV}$ relation has been quantified in detail in the literature, across a wide range of redshifts from the local Universe to $z\approx8$ \citep{meurer1999,calzetti2000,capak2015,barisic2017,fudamoto2020,schouws2022,bowler2024,hamed2024,mondal2024,villanueva2024,bakx2025,fisher2026} or even beyond through stacking \citep{bouwens2020,ciesla2025}. However, $\mathrm{IRX}$ measurements at high redshift are particularly uncertain due to typically poorly constrained dust SEDs, and the resulting large systematic uncertainties on $L_\mathrm{IR}$. At the same time, the common assumption of a fixed dust temperature at high-$z$ may artificially reduce scatter, or yield systematic offsets from the true relation. Moreover, the precise position of a galaxy in the $\mathrm{IRX}-\beta_\mathrm{UV}$ plane depends not only on its dust properties such as the level of reddening and the attenuation law, but also on its stellar age and the star-to-dust geometry \citep[e.g.,][]{faisst2017,popping2017_irxbeta,vijayan2024,nakazato2026}. Given the complex interplay of all these factors in setting the IRX-$\beta_\mathrm{UV}$ relation, our understanding of far-infrared dust properties at high redshift has consequently remained rather limited, with various studies finding significant differences in the implied dust law even at similar epochs and for ostensibly similar galaxy samples \citep[e.g.,][]{capak2015,fudamoto2020,schouws2022,bowler2024}.

While limited to a single galaxy, the Big Three Dragons benefits from a well-constrained dust SED, as well as \textit{JWST}/NIRCam imaging yielding accurate measurements of its UV continuum slope \citep{sugahara2024}. The $\mathrm{IRX}-\beta_\mathrm{UV}$ relation of the system and its two components is presented in Figure \ref{fig:irx_beta}, and compared to additional single- and multi-band-detected galaxies in the literature. Intriguingly, the Eastern component -- which \citet{sugahara2024} find to be highly dust-reddened with $\beta_\mathrm{UV} = -1.18^{+0.15}_{-0.16}$ -- falls in between the canonical \citet{calzetti2000} and SMC \citep{gordon2003} relations (anchored to an intrinsic $\beta_\mathrm{UV,0} = -2.62$ based on \citealt{reddy2018}), and is formally consistent with both within the uncertainties. The Western component, on the other hand, falls $\gtrsim1\,\mathrm{dex}$ above the \citet{calzetti2000} relation owing to its high infrared luminosity, yet apparently modest level of reddening in NIRCam. The position of the global system on the $\mathrm{IRX}-\beta_\mathrm{UV}$ plane lies in between that of its two individual clumps, and is formally consistent with the \citet{calzetti2000} curve, while falling above the SMC relation. The global IRX value decreases slightly when using the fixed $\beta_\mathrm{IR} = 2.0$ fit, placing the system closer to the \citet{calzetti2000} relation (Appendix \ref{app:fixed_beta2}).

As discussed, deviations from the $\mathrm{IRX}-\beta_\mathrm{UV}$ relation can be due to a variety of effects, such as stellar age (reddens $\beta_\mathrm{UV}$), steeper dust laws than the \citet{calzetti2000} relation (reduces $\mathrm{IRX}$ at fixed $\beta_\mathrm{UV}$), and geometrical effects. Based on \citet{sugahara2024}, the Big Three Dragons is a particularly young starburst ($t\sim35\,\mathrm{Myr}$; averaged across the entire system), which suggests that any reddening of its UV slope can be attributed fully to dust. For the Eastern clump of the Big Three Dragons, its position in the $\mathrm{IRX}-\beta_\mathrm{UV}$ plane is thus readily explained by a simple foreground dust screen with an SMC- or Calzetti-like attenuation curve. This is consistent with recent work by \citet{fisher2026} for similarly UV-luminous galaxies at $z\sim7$, who indeed find a typical attenuation curve in between the \citet{calzetti2000} and SMC relations, although we caveat that their study relies mostly on single-band dust temperature estimates. 

The high $\mathrm{IRX}$ yet relatively blue UV slope for the Western component, on the other hand, is likely the result of a more complex geometry than a foreground screen. In the models of \citet[][see also \citealt{faisst2017}]{popping2017_irxbeta}, this combination can be reproduced by a `dust screen with holes', whereby a small fraction of the UV luminosity escapes almost completely unattenuated, yielding a modest $L_\mathrm{UV}$ with a blue continuum slope. Most of the UV emission, however, is nearly completely obscured, resulting in a large infrared luminosity and thus a high $\mathrm{IRX}$. Similar patchy dust has been observed in high-resolution ALMA observations of Y1 \citep{tamura2023,bakx2025}, which thus suggests this geometry may be common in hot, dusty and possibly merging galaxies.

\subsubsection{The $T_\mathrm{dust} - \Sigma_\mathrm{SFR}$ relation}
\label{sec:Tdust_Sigma}

\begin{figure}
	\begin{center}
		\includegraphics[width=0.95\columnwidth]{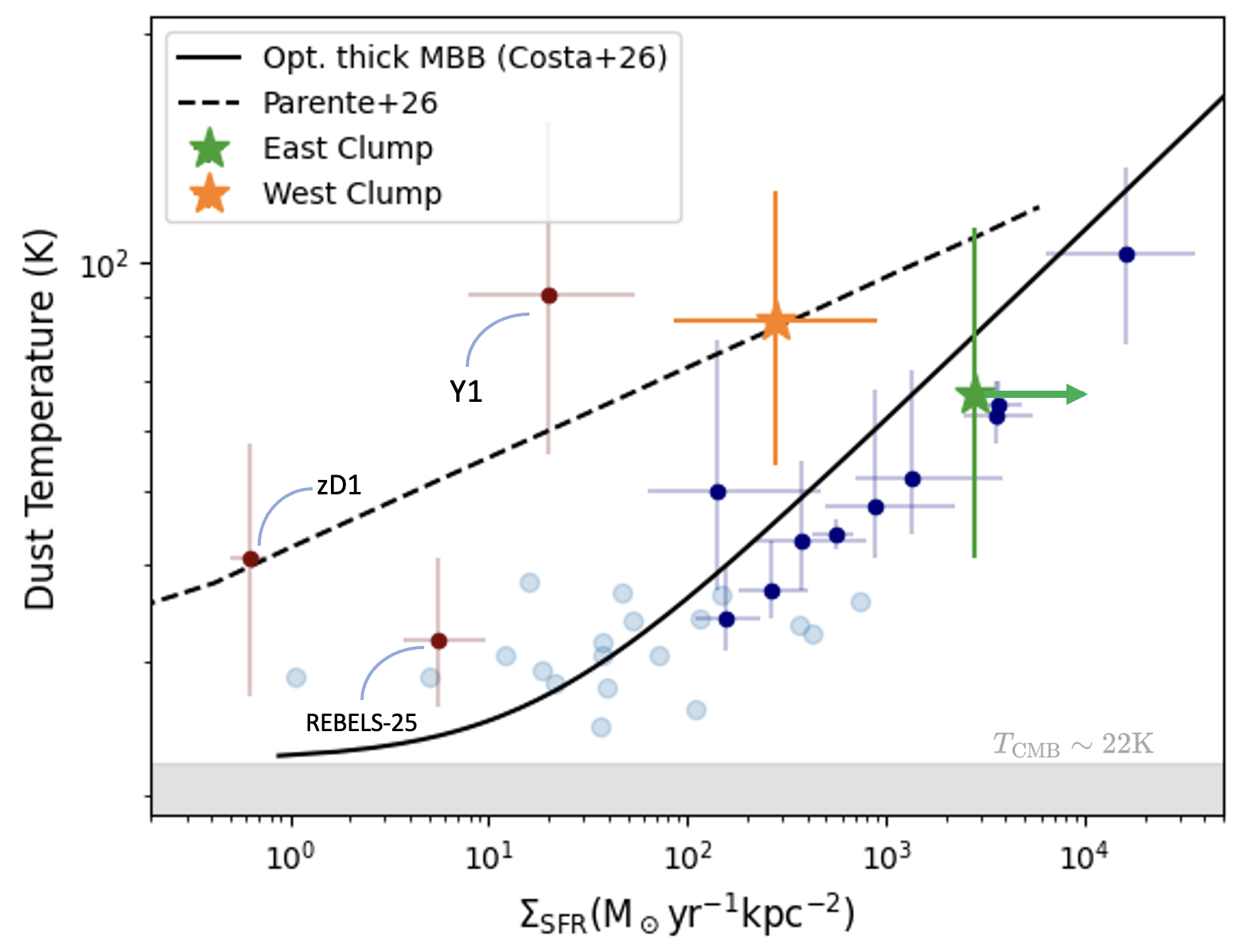}
		\caption{Dust temperature versus SFR surface density for the East and West clumps of the Big Three Dragons, compared with representative high redshift galaxies. The \textit{navy blue} points represent quasar host galaxies at $z \sim 6.5$ from \citet{costa2026}, while the $z=7.1 - 8.3$ Lyman-break galaxies REBELS-25, Y1, and zD1 are labeled and shown in \textit{dark red}. Dusty star-forming galaxies (DSFGs) from \citet{bakx2024} are plotted in \textit{light blue}. The \textit{solid line} (calculated following \citealt{costa2026}) represents the theoretical optically thick limit for a modified blackbody, assuming $T_\mathrm{CMB}(z=7.15)=22\,$K. The \textit{dashed line} shows the broken power law fit for the $\Sigma_\mathrm{SFR}-T_\mathrm{dust}$ relation from \citet{parente2026}. The Western clump agrees with this relation, while the dust temperature of the Eastern one could plausibly be underestimated due to the presence of optically thick dust.}
        \label{fig:sigsfr}
	\end{center}
\end{figure}

\noindent Finally, we plot the dust temperatures of the two LBGs making up the Big Three Dragons versus SFR surface density, $\Sigma_\mathrm{SFR}$, in Figure \ref{fig:sigsfr}. As discussed in further detail below, this comparison provides qualitative insight into whether the dust in the two LBGs may be optically thick, as well as into the possible presence of an AGN within them.

As none of the multi-band ALMA continuum observations robustly resolve the two LBGs, we compute their clump-based $\Sigma_\mathrm{SFR}$ from the NIRCam rest-UV sizes reported by \citet[][see also \citealt{sugahara2024}]{prieto-jimenez2025}. They determine effective radii of $<63\,\mathrm{pc}$ and $\sim350\,\mathrm{pc}$ for the Eastern and Western clumps, respectively, with the former remaining unresolved even in NIRCam/F115W, and thus constituting an upper limit. We note that \citet{mitsuhashi2024_cristal} find dust to be $\sim2\times$ more extended than the UV emission in $z\sim5$ main-sequence galaxies observed as part of CRISTAL \citep{herrera-camus2025}; if also the case for the Big Three Dragons, this would decrease the inferred star formation rate surface densities by a factor of $\sim4\times$. We furthermore note that the Eastern clump is surrounded by a fainter, tidal-like structure that is more spatially extended \citep[][]{sugahara2024}. If a large fraction of the dust emission in the Eastern component emanates from this extended tail, its SFR surface density may be further overestimated. New, high-resolution Band 8 observations of the Big Three Dragons, to be presented in Arai et al.\ (in preparation), are set to provide additional insight into the dust continuum sizes of the two clumps.

In our $T_\mathrm{dust}-\Sigma_\mathrm{SFR}$ comparison, we also include the $z\sim6.5$ quasar host galaxies recently compiled by \citet{costa2026}, the DSFG sample of \citet{bakx2024_angels} which benefits from high-resolution ALMA observations in Bands 3 through 8, and the individual Lyman-break galaxies REBELS-25, Y1, and zD1. For the \citet{costa2026} quasars, where sizes were obtained from 2D Gaussian fitting, we convert this into an effective radius as $r_\mathrm{eff} = 0.5\sqrt{\theta_\mathrm{major} \times \theta_\mathrm{minor}}$. We then derive the star formation rate surface density as $\Sigma_\mathrm{SFR} = \mathrm{SFR}_\mathrm{IR} / (2\pi r_\mathrm{eff}^2)$ where the factor $1/2$ accounts for the fact that approximately half of the star formation is expected to be concentrated within $<r_\mathrm{eff}$. For REBELS-25, we adopt the total $\mathrm{SFR}_\mathrm{UV+IR}$ from \citet{fisher2026}, and the effective radius from Rowland et al.\ (in preparation) based on spatially resolved dust observations in ALMA Bands 6 and 8. For zD1, we adopt the total $\mathrm{SFR}_\mathrm{UV+IR}$ from \citet{akins2022} and size from \citet[][see also \citealt{posses2025}]{knudsen2025}, though we caution the latter is based on \ciil{} emission which is likely more extended than the dust continuum. All values for Y1 were taken from \citet{bakx2025}. In cases where no uncertainties on the radii were reported, we assume a fixed 10\% uncertainty.

The solid line in Figure \ref{fig:sigsfr} corresponds to the relation for fully optically thick dust emission (i.e., $e^{-\tau_{\nu}} \sim 0$) against the CMB at $z=7.15$, and is determined following \citet{costa2026}. This relation corresponds to perfect blackbody emission, and thus represents the theoretical upper limit on $\Sigma_\mathrm{SFR}$ at a given dust temperature. The generally compact quasar host galaxies from \citet{costa2026} lie just beyond this limit, which is consistent with the likely presence of hot, optically thick dust surrounding their AGN.\footnote{We note that, for their quasar sample, \citet{costa2026} define the size as the radius encompassing $\sim99\%$ of the continuum emission. Overall, the inferred SFR surface densities in their work are therefore $\sim3\times$ smaller than the ones derived here, which are determined using the effective radius for consistency with the Big Three Dragons and other literature samples.} The Eastern clump of the Big Three Dragons, in line with its compact size, appears to have similar properties as these quasar host galaxies, suggesting that it may contain optically thick dust or even host an AGN, although the uncertainties on $T_\mathrm{dust}$ and $\Sigma_\mathrm{SFR}$ (through the unknown dust continuum size) remain large. We note that \textit{JWST} spectroscopy of the Big Three Dragons reveals no concrete evidence that it hosts a (luminous) AGN \citep{prieto-jimenez2025,jones2024}, though based on its high implied SFR surface density, \citet{sugahara2024} have previously argued that the bright, compact emission seen in the Eastern clump could be partially AGN-powered.

The presence of optically thick dust in the Eastern clump would imply a warmer intrinsic dust temperature than that inferred from our fiducial optically thin model ($T_d = 67_{-26}^{+44}\,\mathrm{K}$), and would thus further reinforce our conclusions that this clump contains hot dust. On the other hand, the presence of optically thick dust would not significantly affect the infrared luminosity of the clump \citep[e.g.,][]{algera2024,sommovigo_algera2025}. If it indeed contains a dust-obscured AGN, however, some of the infrared emission may be powered by the central engine, which could lead to a (modest) overestimation of its star formation rate \citep[see also e.g.,][]{meyer2025,silverman2025}.

The Western clump of the Big Three Dragons, on the other hand, is more spatially extended. Its dust temperature agrees well with the broken power law fit from \citet[][dashed line in the figure]{parente2026}, who find that an elevated $\Sigma_\mathrm{SFR}$ is the primary driver of the warm dust temperatures seen at high redshift. For this component, we thus find no strong evidence that the dust is optically thick.

\subsection{Implications for the obscured SFRD at $z\approx7$}
\label{sec:discussionInterpretation}

\begin{figure*}
    \centering
    \includegraphics[width=0.75\textwidth]{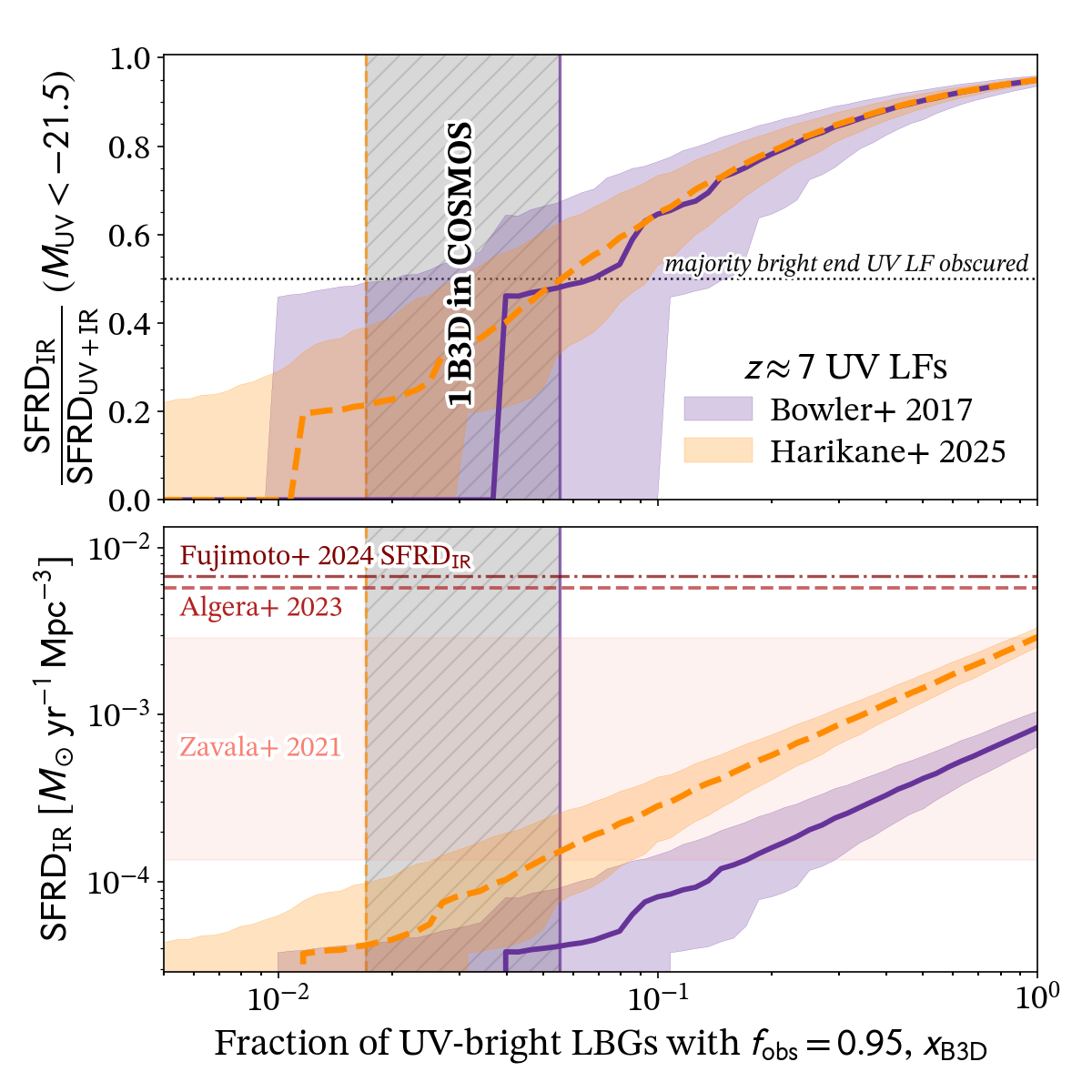}
    \caption{\textit{Top:} the obscured fraction across the full UV-bright ($M_\mathrm{UV}<-21.5$) LBG population at $7 < z < 7.5$, as a function of the incidence rate of Big Three Dragons-like (B3D-like) systems, $x_\mathrm{B3D}$, defined such that for $x_\mathrm{B3D}=1$ all UV-luminous LBGs with $M_\mathrm{UV}<-21.5$ have an obscured fraction of $f_\mathrm{obs}=0.95$. We show the results for the two UV LFs from \citet[][solid purple; shading represents the $1\sigma$ uncertainty]{bowler2017} and \citet[][dashed orange]{harikane2025_uvlf}. \textit{Bottom:} the corresponding $\mathrm{SFRD}_\mathrm{IR}$ contributed by B3D-like galaxies, together with the independent $z\sim7$ IR SFRD constraints from \citet{zavala2021}, \citet{algera2023}, and \citet{fujimoto2023}. In both panels, the grey band corresponds to the scenario in which the Big Three Dragons is the only highly obscured LBG in COSMOS at this epoch. This band alone already implies that half of all star formation in the UV-bright population is obscured for the \citet{bowler2017} LF, and $\sim20\%$ for the \citet{harikane2025_uvlf} one. The total implied $\mathrm{SFRD}_\mathrm{IR}$ due to B3D-like galaxies is considerable if they are reasonably common ($x_\mathrm{B3D} \gtrsim 0.1 - 0.2$), matching the lower end of the \citet{zavala2021} prediction across the full $z\sim7$ galaxy population. Overall, this highlights that hot, highly obscured galaxies can contribute significantly to the star formation budget of the UV-bright galaxy population.}
    \label{fig:fracSFRD}
\end{figure*}

The high global IR luminosity of the $z=7.15$ Big Three Dragons implies a high obscured star formation rate ($\mathrm{SFR}_\mathrm{IR} = 251_{-153}^{+418}\,M_\odot\,\mathrm{yr}^{-1}$) and obscured fraction ($f_\mathrm{obs} = 0.94_{-0.09}^{+0.04}$). If such hot, highly obscured systems are common in the early Universe, this could have important implications for the obscured SFRD at $z>7$. We quantify this in what follows, utilizing the fact that the Big Three Dragons is drawn from a well-defined survey volume -- namely, the $1.65\,\mathrm{deg}^2$ COSMOS field \citep[e.g.,][]{bowler2012,bowler2014}.

As a starting point, we make the clearly wrong assumption that the Big Three Dragons alone accounts for \textit{all} obscured star formation at $7 < z < 7.5$ in COSMOS. This yields a strict lower limit to the overall obscured SFRD at this epoch, given that additional dusty $z>7$ galaxies in the field are known to exist \citep[e.g.,][]{inami2022,schouws2022}.\footnote{Conversely, if galaxies like the Big Three Dragons are especially rare, and we were therefore lucky to find even one in the degree-scale COSMOS field, the implied $\mathrm{SFRD}_\mathrm{IR}$ could still be an overestimate; we discuss this further below.} By itself, the Big Three Dragons would account for $\mathrm{SFRD}_\mathrm{IR} = 4.0_{-2.4}^{+6.8} \times 10^{-5}\,M_\odot\,\mathrm{yr}^{-1}\,\mathrm{Mpc}^{-3}$, which corresponds to a negligible obscured fraction of $f_\mathrm{obs} = 3.1_{-1.9}^{+5.3}\times10^{-3}$ relative to the unobscured SFRD from \citet{harikane2025_uvlf}, integrated down to $M_\mathrm{UV} = -13$. This framing -- one obscured, UV-bright system compared to the full unobscured star-forming population -- is of course extremely conservative, and in what follows we instead ask a more informative question: what fraction of the total SFR budget of the UV-bright population is obscured if systems like the Big Three Dragons are not unique? \\

We first recall that the Big Three Dragons was originally identified as a $z\approx7$ UV-luminous LBG \citep[$M_\mathrm{UV} = -22.4$;][]{bowler2014}, motivating a comparison against other UV-luminous systems in particular. While there is growing evidence for an additional highly obscured $z\gtrsim6$ galaxy population that is not necessarily UV-bright \citep[e.g.,][]{fudamoto2021,endsley2023,bakx2024,vanleeuwen2024,bing2025,zavala2026}, the number densities and IR luminosities of this population are currently especially uncertain. For the purpose of estimating the effect of UV-bright yet obscured `Big Three Dragons-like' (henceforth, B3D-like) galaxies on the obscured SFRD, we will therefore focus solely on UV-luminous systems.

We first outline the core idea of our approach, while discussing the details in what follows. Briefly, we draw UV-bright galaxies from the UV luminosity function (LF) and designate a fraction $0 \leq x_\mathrm{B3D} \leq 1$ of them as B3D-like, assigning them an obscured fraction of $f_\mathrm{obs} = 0.95$. For the remainder, we adopt an obscured fraction of $f_\mathrm{obs} = 0$. This choice is deliberately conservative, and is meant to isolate the excess obscured SFRD contributed specifically by the B3D-like galaxy population, rather than folding in additional assumptions about dust obscuration among ordinary UV-selected galaxies. We perform this assignment in a Monte Carlo approach, such that the precise number of galaxies assigned a high obscured fraction can vary stochastically between realizations. By summing up the total $\mathrm{SFR}_\mathrm{IR}$ from these UV-bright sources and using the known area across which they are selected, we infer both the global obscured fraction across the UV-bright population and the corresponding $\mathrm{SFRD}_\mathrm{IR}$ as a function of the incidence rate $x_\mathrm{B3D}$.

In practice, we adopt the UV LFs from \citet{bowler2017} and \citet{harikane2025_uvlf}, parameterized as Schechter functions, and stochastically draw $7 < z < 7.5$ galaxies with $M_\mathrm{UV} < - 21.5$ across an area of $1.65\,\mathrm{deg}^2$ in a Monte Carlo approach, assuming Poisson statistics. We adopt two different LFs to account for variance in the number of predicted UV-luminous galaxies. The adopted magnitude limit is chosen to coincide with the selection criteria of the REBELS survey \citep{bouwens2022}, which has successfully identified several dusty galaxies at $z > 7$ \citep{inami2022}. Moreover, this limit is typical for ground-based studies of the UV LF conducted across degree-scale fields \citep[e.g.,][]{bowler2017,varadaraj2023}. The UV LFs from \citet{bowler2017} and \citet{harikane2025_uvlf} yield estimates of $N_\mathrm{LBG}\sim18$ and $N_\mathrm{LBG}\sim59$ UV-bright LBGs at $7 < z < 7.5$ across the COSMOS field, respectively. If the Big Three Dragons is thus the only UV-bright, heavily obscured galaxy in COSMOS at this epoch, the corresponding incidence rate is $x_\mathrm{B3D} = 1 / N_\mathrm{LBG} \sim 0.02 - 0.06$. \\

We show the effect of B3D-like galaxies on obscured cosmic star formation in Figure \ref{fig:fracSFRD}. The top panel shows the fraction of the total SFR budget across the UV-bright population that is dust-obscured, denoted $F_\mathrm{obs} = \mathrm{SFRD}_\mathrm{IR} / \mathrm{SFRD}_\mathrm{UV+IR}\,(M_\mathrm{UV} < -21.5)$, as a function of $x_\mathrm{B3D}$. The bottom panel shows the corresponding obscured $\mathrm{SFRD}_\mathrm{IR}$ itself, together with independent literature constraints at $z\sim7$ determined across the full galaxy population \citep{zavala2021,algera2023,fujimoto2023}. In both panels, the grey hatched band marks the scenario in which the Big Three Dragons is the only B3D-like galaxy in the COSMOS field at this epoch. Given the \citet{bowler2017} LF, a single B3D-like galaxy contributes as much star formation as the full UV-bright population across COSMOS, i.e. $F_\mathrm{obs} \approx 0.5$,\footnote{This follows readily from the fact that a galaxy with $f_\mathrm{obs} = 0.95$ has $\mathrm{SFR}_\mathrm{IR} / \mathrm{SFR}_\mathrm{UV} = 19$, which is approximately the number of UV-bright LBGs predicted across COSMOS by \citet[][$N_\mathrm{LBG}\approx18$]{bowler2017}.} while for the \citet{harikane2025_uvlf} LF, one Big Three Dragons corresponds to $F_\mathrm{obs} \approx 0.2$. More generally, for an incidence rate of $x_\mathrm{B3D} \gtrsim 0.1$, both the \citet{bowler2017} and \citet{harikane2025_uvlf} LFs predict that $F_\mathrm{obs} \gtrsim 0.5$, i.e., obscured star formation dominates the total SFR budget of UV-luminous LBGs at this epoch, even under fairly conservative assumptions about the rarity of such highly obscured systems.

The bottom panel of Figure \ref{fig:fracSFRD} shows the total $\mathrm{SFRD}_\mathrm{IR}$ contributed by B3D-like galaxies, and reveals that even for a modest incidence of $x_\mathrm{B3D} \gtrsim 0.1 - 0.2$, the total obscured SFRD of this population roughly matches the lower range of the obscured $z\approx7$ SFRD inferred across the \textit{full} galaxy population by \citet{zavala2021}. However, even in the extreme scenario where $x_\mathrm{B3D} = 1$ and thus every UV-luminous LBG is in fact highly obscured with $f_\mathrm{obs} = 0.95$, the resulting $\mathrm{SFRD}_\mathrm{IR}$ remains well below the constraints from \citet{algera2023} and \citet{fujimoto2023}. This is not surprising given that -- by construction -- the exercise presented here yields a strict lower limit on the obscured $\mathrm{SFRD}$ at $z\approx7$, as no obscured star formation in the UV-faint population has been accounted for.

We proceed by estimating the contribution of B3D-like galaxies relative to the total cosmic SFRD at $z\approx7$, by integrating the \citet{bowler2017} and \citet{harikane2025_uvlf} LFs down to $M_\mathrm{UV} = -13$. In such a cosmological context, we find the contribution of such UV-bright, highly obscured galaxies to be only minor; in the `single-B3D in COSMOS' scenario, the Big Three Dragons accounts for $<1\%$ of the total SFRD, rising to just a few percent for $x_\mathrm{B3D} = 0.2$. This, in turn, reflects the fact that the bright end of the UV LF contributes only a small fraction to the total dust-unobscured SFRD \citep[e.g.,][]{bouwens2022_uvlf,atek2026}. 

Estimating the value of $x_\mathrm{B3D}$ empirically is difficult at this stage, given that currently only four $z>7$ LBGs have robustly measured infrared luminosities based on combined observations of the peak and Rayleigh-Jeans tail of the dust SED \citep[e.g.,][this work]{bakx2021,bakx2025,algera2024b}. Interestingly, however, the two $z>7$ galaxies with known hot dust temperatures -- the Big Three Dragons and Y1 -- are both likely merging galaxies \citep{harshan2024,sugahara2024}, while the $z=7.31$ galaxy REBELS-25 -- which has a particularly low $T_d$ \citep{algera2024b} -- instead is known to be a dynamically cold, rotating disk \citep{rowland2024}. This could suggest a connection between a merger-induced starburst and a warm dust reservoir, possibly due to the starburst both rapidly producing and simultaneously heating the dust. Given that merger rates likely increase with redshift \citep[e.g.,][]{romano2021,duan2025,puskas2025}, this could suggest that hot, highly obscured starbursts account for an increasingly important fraction of the obscured SFRD at early cosmic times -- consistent with the picture above, in which even a modest incidence of such systems already dominates the SFR budget of the UV-bright population. While current sample sizes remain small, future efforts into measuring the dust properties of known high-redshift mergers may shed further light on this possible connection.

\section{Conclusions}
\label{sec:conclusions}

We have presented new ALMA Band 9 ($\lambda_\mathrm{rest} \approx 55\,\mu\mathrm{m}$) and Band 4 ($\lambda_\mathrm{rest} \approx 270\,\mu\mathrm{m}$) observations towards Big Three Dragons, a UV-luminous yet dusty pair of merging Lyman-break galaxies at $z=7.15$. The system is detected at $3.5\sigma$ in the unresolved Band 4 observations, and the higher resolution Band 9 data detect both individual LBGs at $3.7\sigma$ and $3.5\sigma$. Combined with previous continuum detections in ALMA Bands 6, 7 and 8 \citep{bowler2018,hashimoto2019,sugahara2021}, we infer the following:

\begin{itemize}
    \setlength{\itemsep}{0.2cm}
    
    \item \textbf{Hot dust}: through an optically thin modified blackbody fit, we infer a hot global dust temperature of $T_d = 78_{-23}^{+35}\,\mathrm{K}$ for the Big Three Dragons (Fig.\ \ref{sed}). The fit also yields a modest dust mass $\log(M_d/M_\odot) = 6.85^{+0.33}_{-0.27}$ and typical dust emissivity index $\beta_\mathrm{IR} = 1.47_{-0.35}^{+0.46}$. The ALMA observations in Bands 6, 8 and 9 moreover spatially resolve the two main components of the Big Three Dragons (East and West), and we find that both are characterized by a similarly hot dust temperature ($T_d \approx 67 - 84\,\mathrm{K}$). This implies the Big Three Dragons is one of the hottest galaxies known at this epoch, placing it above most theoretical $T_d-z$ relations at $z\approx7$ (Fig.\ \ref{fig:tdust}).

    \item \textbf{Highly obscured:} the presence of hot dust in the Big Three Dragons implies a high $\log(L_\mathrm{IR}/L_\odot) = 12.32^{+0.43}_{-0.41}$ for the combined system, making it one of the most IR-luminous galaxies known at $z>7$ (Fig.\ \ref{fig:lir_redshift}). Its total star formation rate is $\mathrm{SFR}_\mathrm{UV+IR} = 267_{-154}^{+419}\,M_\odot\,\mathrm{yr}^{-1}$, of which a large fraction of $f_\mathrm{obs} = 0.94^{+0.04}_{-0.09}$ is dust-obscured. This places the Big Three Dragons well above the $f_\mathrm{obs} - M_\star$ relation established at lower redshifts (Fig.\ \ref{fig:fobs}).

    \item \textbf{IRX-$\mathbf{\beta_\text{UV}}$:} the Eastern clump of the Big Three Dragons falls in between the canonical Calzetti and SMC $\mathrm{IRX}-\beta_\mathrm{UV}$ relations, while the Western clump lies $>1\,\mathrm{dex}$ above the \citet{calzetti2000} curve (Fig.\ \ref{fig:irx_beta}). For the latter, this is likely due to patchy dust obscuration in the system, in which most of the UV emission is fully obscured and thus re-emitted in the far-infrared, but a small fraction is emitted without being strongly reddened.

    \item \textbf{$\mathbf{T_\text{dust} - \Sigma_\text{SFR}}$}: the warm dust temperature of the Western LBG is likely driven by its high SFR surface density, based on its good agreement with the $T_\mathrm{dust} - \Sigma_\mathrm{SFR}$ relation from \citet[][Fig.\ \ref{fig:sigsfr}]{parente2026}. The compact Eastern LBG, on the other hand, falls below the relation, despite its seemingly especially high $\Sigma_\mathrm{SFR} \gtrsim 3\times10^{3}\,M_\odot\,\mathrm{yr}^{-1}\,\mathrm{kpc}^{-2}$. This suggests the possible presence of optically thick dust -- causing the currently inferred warm dust temperature to be a lower limit -- and/or a hidden AGN causing the SFR surface density to be overestimated.

    \item \textbf{Contribution to the SFRD:} even if the Big Three Dragons were the \textit{only} highly dust-obscured ($f_\mathrm{obs} = 0.95$) galaxy in COSMOS at $7 < z < 7.5$, it alone accounts for $20 - 50\%$ of all star formation occurring at the bright end of the UV luminosity function ($M_\mathrm{UV} < -21.5$) at this epoch. Compared to the total cosmic star formation rate density, however, the contribution of heavily-obscured, UV-luminous galaxies is likely minor ($\lesssim5\,\%$) unless they turn out to be particularly common ($>20\%$ of all UV-bright LBGs).

    
\end{itemize}

Altogether, this study paints the Big Three Dragons as a uniquely luminous $z\approx7$ galaxy, which further highlights the importance of dust-obscured star formation well into the epoch of reionization. Despite the current multi-band sampling of the Big Three Dragons in ALMA Bands 3 through 9, further follow-up in Band 10 is necessary to better constrain its dust temperature and obscured SFR. While the Big Three Dragons itself is uniquely bright, the upcoming ALMA Wideband Sensitivity Upgrade \citep{carpenter2022} will make such dedicated multi-frequency dust continuum observations feasible for a much larger sample of high-redshift galaxies, and is thereby set to greatly increase our understanding of both the production and properties of dust at early cosmic times.

\section*{Acknowledgments}

This paper makes use of the following ALMA data: \\ ADS/JAO.ALMA\#2016.1.00954.S, ADS/JAO.ALMA\#2017.1.00190.S, ADS/JAO.ALMA\#2018.1.01673.S, ADS/JAO.ALMA\#2019.1.01491.S, ADS/JAO.ALMA\#2023.1.01033.S. 

ALMA is a partnership of ESO (representing its member states), NSF (USA) and NINS (Japan), together with NRC (Canada), MOST and ASIAA (Taiwan), and KASI (Republic of Korea), in cooperation with the Republic of Chile. The Joint ALMA Observatory is operated by ESO, AUI/NRAO and NAOJ. HSBA gratefully acknowledges support from Academia Sinica through grant AS-PD-1141-M01-2. AKI acknowledges support from KAKENHI No. 26H02069. YS acknowledges support from JSPS KAKENHI Grant Number JP26K17200. RHC thanks the Max Planck Society for support under the Partner Group project "The Baryon Cycle in Galaxies" between the Max Planck for Extraterrestrial Physics and the Universidad de Concepción. RHC also gratefully acknowledges financial support from ANID - MILENIO - NCN2024\_112 and ANID BASAL FB210003. YF acknowledges support from JSPS KAKENHI Grant Numbers JP22K21349 and JP23K13149. MA is supported by FONDECYT grant number 1252054, and gratefully acknowledges support from ANID Basal Project FB210003, ANID MILENIO NCN2024\_112 and ANID + Vinculaci\'on Internacional + FOVI250261.

\section*{Data Availability}

All data utilized in this work is available upon reasonable request to the corresponding authors.



\bibliographystyle{mnras}
\bibliography{main} 


\appendix

\section{Tapered Band 9 data}
\label{app:taper}

\begin{figure}
	\begin{center}
		\includegraphics[width=0.9\columnwidth]{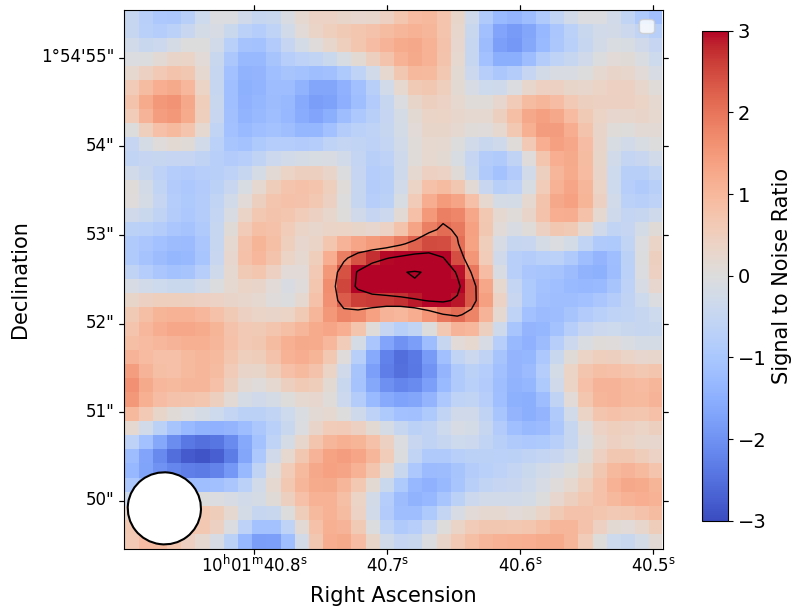}
		\caption{The ALMA Band 9 image tapered to a coarser resolution of 0.8 arcsec, where it is detected at $4~\sigma$ significance.} 
            \label{fig:taper}
	\end{center}
\end{figure}

The Band 9 observations spatially resolve the dust emission in the Big Three Dragons (Figure \ref{fig:cutouts}), although as a result both clumps are detected at only a modest significance ($3.5 - 3.7\sigma$). To improve the S/N at the expense of angular resolution, we also create tapered images. As an example, we show the Band 9 cutout tapered to $0.8''$ resolution in Figure \ref{fig:taper}, where the galaxy is detected at $\sim4\sigma$. We moreover note that the global flux measurements of the Big Three Dragons are consistent between the naturally-weighted and tapered images.

\section{Flux measurements}
\label{app:fluxes}
\begin{figure}
    \centering

    \begin{subfigure}{0.9\columnwidth}
        \centering
        \includegraphics[width=\linewidth]{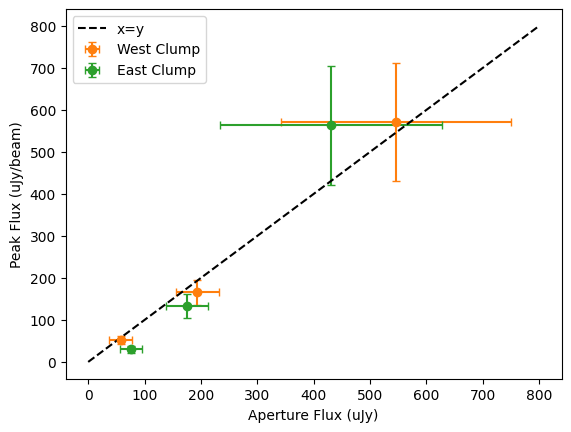}
        \label{fig:peakflux}
    \end{subfigure}
    \hfill
    \begin{subfigure}{0.9\columnwidth}
        \centering
        \includegraphics[width=\linewidth]{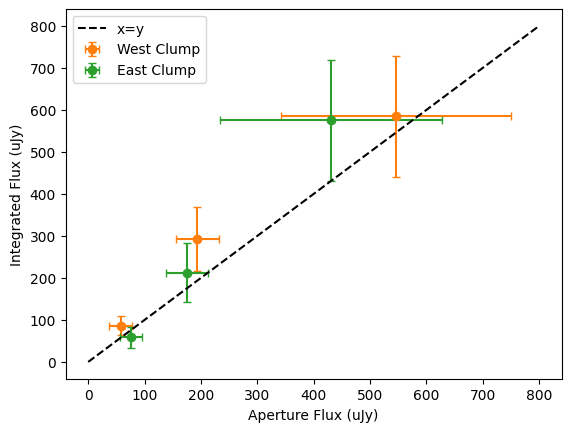}
        \label{fig:intflux}
    \end{subfigure}

    \caption{Comparison of aperture photometry with CASA \texttt{imfit} flux measurements for Bands 6, 8 and 9.}
    \label{fig:imfit}
\end{figure}

We compare the aperture flux measurements for the resolved clumps with fluxes derived from CASA \texttt{imfit}, where we perform a two-component, 2D-Gaussian fit to the Band 6, 8 and 9 data. For the Band 6 and Band 8 fits, the source positions and size parameters are left free. However, this approach did not work for the Band 9 image due to its low signal-to-noise ratio. Assuming the clumps can be treated as point sources, we therefore fixed the major and minor axes of the Band 9 Gaussian components to match the shape of the Band 9 beam, and confirm this yields a good fit. Figure \ref{fig:imfit} shows that both the peak (top panel) and integrated (bottom) flux densities from the 2D Gaussian fits agree well with the aperture measurements, and as a result the inferred dust properties of the Big Three Dragons do not significantly change if these Gaussian fluxes are adopted.

\begin{table*}
    \centering
    \label{tab:gaussian_fluxes}
    \begin{tabular}{lcccccc}
        \hline
        Component & Integrated Flux ($\mu$Jy) & Peak ($\mu$Jy\,beam$^{-1}$) & Major ($''$) & Minor ($''$) & PA ($^\circ$) \\
        \hline
        Band 6 West & $86 \pm 23$ & $53 \pm 9$ & 0.345 & 0.307 & 41.7 \\
        Band 6 East & $58 \pm 24$ & $30 \pm 8$ & 0.544 & 0.230 & 110.2 \\
        \hline
        Band 8 West & $292 \pm 76$ & $165 \pm 30$ & 0.626 & 0.516 & 99.4 \\
        Band 8 East & $213 \pm 70$ & $134 \pm 29$ & 0.648 & 0.448 & 144.6 \\
        \hline
        Band 9 West & $584 \pm 144$ & $571 \pm 141$ & 0.450 & 0.390 & 69.2 \\
        Band 9 East & $576 \pm 144$ & $563 \pm 141$ & 0.450 & 0.390 & 83.3 \\
        \hline
    \end{tabular}
    \caption{Flux measurements from two-component Gaussian fits using CASA \texttt{imfit}.}
\end{table*}

\section{Corner plot}
\label{app:corner}

\begin{figure}
	\begin{center}
		\includegraphics[width=0.95\columnwidth]{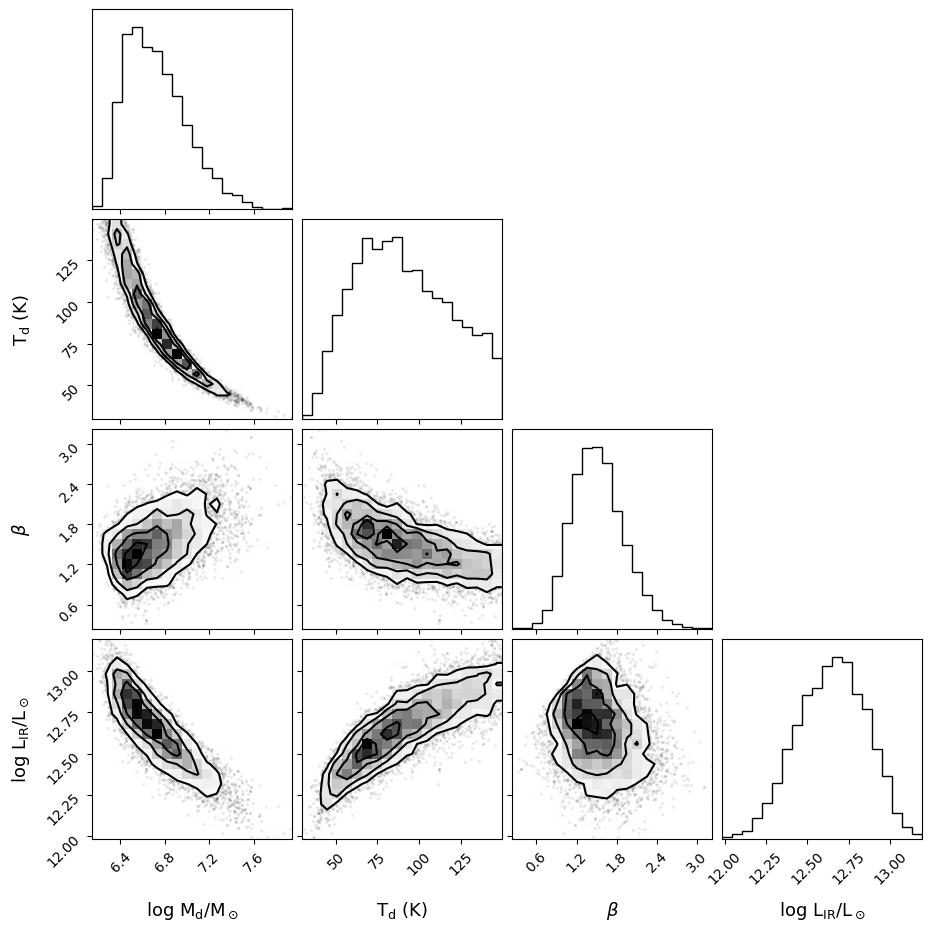}
		\caption{Corner plot showing the posterior parameter distributions from our MBB fit to the global dust SED of the Big Three Dragons. The derived infrared luminosity ($L_\mathrm{IR}$), computed for each posterior sample, is also shown to highlight its degeneracies with the fitted parameters.}
            \label{fig:corner}
	\end{center}
\end{figure}

We show a corner plot \citep{foreman-mackey2016} for the global MBB fit to the Big Three Dragons in Figure \ref{fig:corner}.

\section{Fit with $\beta_\mathrm{IR} = 2$}
\label{app:fixed_beta2}

\begin{figure}
	\begin{center}
        \includegraphics[width=0.95\columnwidth]{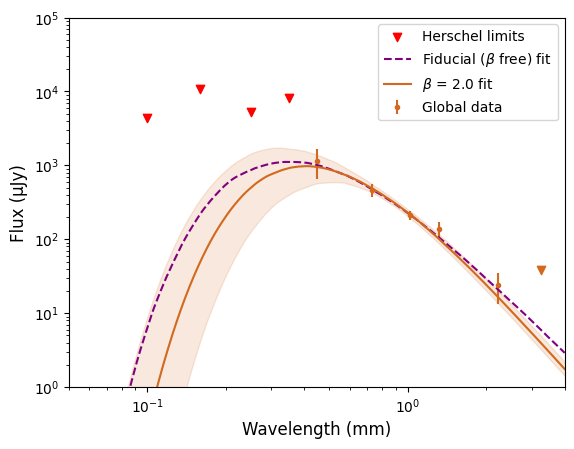}
		\caption{The global MBB fit with the $\beta_\mathrm{IR}$ parameter fixed at 2.0 is shown in \textit{brown}. The fiducial fit, which adopts a gaussian prior on $\beta_\mathrm{IR}$ and yields $\beta_\mathrm{IR} \sim 1.5$, is shown as a dashed line for comparison. The $\beta_\mathrm{IR} = 2.0$ fit yields a lower median dust temperature of 57 K, which is however consistent within the uncertainties with our fiducial result.} 
        \label{fig:beta2}
	\end{center}
\end{figure}

\begin{table}
\begin{tabular}{ll} \hline
Parameter & Value \\ \hline
$\log(M_{\text{d}}/M_\odot)$ & $7.10^{+0.31}_{-0.26}$ \\ 
$T_d$ [K] & $57^{+15}_{-13}$ \\ 
$\log(L_{\mathrm{IR}}/L_\odot)$ & $12.05^{+0.40}_{-0.31}$ \\ 
SFR [$M_\odot$/yr] & $134^{+200}_{-68}$ \\ 
$f_{\mathrm{obs}}$ & $0.89^{+0.05}_{-0.14}$ \\ 
$\log$ IRX & $0.98^{+0.41}_{-0.27}$ \\ \hline
\end{tabular}

\caption{Physical properties of the global system derived from SED fitting with $\beta_\mathrm{IR}$ fixed at 2.0.}
\label{tab:sed_2.0}
\end{table}

In the literature, is typical to assume a fixed dust emissivity index $\beta_\mathrm{IR} = 2.0$ in the absence of observations covering the Rayleigh-Jeans tail of the dust emission \citep[e.g.,][]{sugahara2021,algera2024}. While the Big Three Dragons is detected in Band 4 at the $\sim3.5\sigma$ level, we provide such a fit for completeness in Figure \ref{fig:beta2}, and present the corresponding measurements in Table \ref{tab:sed_2.0}. Given the known anti-correlation between $\beta_\mathrm{IR}$ and dust temperature \citep[e.g.,][]{dacunha2021}, the fit with $\beta_\mathrm{IR} = 2.0$ yields a lower dust temperature (by $\sim20\,\mathrm{K}$) compared to our fiducial fit where we infer $\beta_\mathrm{IR} \approx 1.5$. Consequently, the infrared luminosity of the system decreases slightly by $\sim0.3\,\mathrm{dex}$, although the Big Three Dragons formally remains a ULIRG with an obscured fraction of $f_\mathrm{obs} \sim 90\%$. Nevertheless, this highlights the importance of robust coverage of the Rayleigh-Jeans tail for accurate measurements of high-$z$ galaxy dust properties \citep[e.g.,][]{dacunha2021,algera2024b}, and additional observations in this regime are therefore needed to further improve constraints on the dust properties of the Big Three Dragons.


\bsp	
\label{lastpage}
\end{document}